\documentclass{aa}  

\usepackage{graphicx}
\usepackage{txfonts}
\usepackage{lipsum}
\usepackage{subcaption}         % necessary for continued figures, example in section 3
\usepackage{lscape}             % to rotate a single page table, example in appendix.
\usepackage{placeins}           % useful with \FloatBarrier, to keep 
\usepackage{booktabs}                         % onecolumn floats from drifting to the next section
\usepackage[]{natbib}
\usepackage{graphicx,color}
\usepackage{txfonts}
\usepackage[colorlinks=true, citecolor=blue]{hyperref}
\usepackage[table]{xcolor}
\usepackage{multirow} %multiple row in tables
\usepackage{adjustbox}
\usepackage{float}
\usepackage{booktabs}
 \usepackage[utf8]{inputenc}
\usepackage[T1]{fontenc}                             
\begin{document}

%%%%%%%%%%%%%%%%%%%%%%%%%%%%%%%%%%%%%%%%
% if you use custom commands in your title,
% ensure to check your title when submitting!
%%%%%%%%%%%%%%%%%%%%%%%%%%%%%%%%%%%%%%%%
   \title{Chemo-dynamical stratification of the Galactic disc using \emph{Gaia}-ESO open clusters.}

   \subtitle{}

%%%%%%%%%%%%%%%%%%%%%%%%%%%%%%%%%%%%%%%%
% Please separate each author with the \and command
%
% Use the \corrauth to provide the corresponding
% author address. It will be automatically inserted as 
% footnote in the PDF output.
%
% Please DO NOT include ORCIDs next to author names.
% Instead, please provide an active address for each coauthor:
% it will be automatically extracted by EDPS editorial system, 
% and co-authors will be be able to authenticate their ORCID.
%
% Only authenticated ORCIDs will be taken into account.
% ORCIDs included here will be removed.
%%%%%%%%%%%%%%%%%%%%%%%%%%%%%%%%%%%%%%%%

   \author{C.O Obasi \inst{1}\corrauth{casmiroluabuchukwuobasi@gmail.com}        % use \corrauth for the corresponding author
        \and M. Gómez\inst{1}\email{}%\fnmsep\thanks{NASA fellow (shows the usage of elements in the author field)}
        \and J.~G. Fern\'andez-Trincado\inst{2}\email{}
        \and D. Minniti\inst{1,3}\email{}
        \and B. Dias\inst{1}\email{}
        \and B. Barbuy\inst{4}\email{}
        \and M.~V. Alonso\inst{5,6}\email{}
        \and and L.~D. Baravalle\inst{5}\email{}
        }

   \institute{Instituto de Astrof\'isica, Depto. de Física y Astronomía, Facultad de Ciencias Exactas, Universidad Andr\'es Bello, Av. Fern\'andez Concha 700, Las Condes, Santiago, Chile.
   \and Centro de investigaci\'on en Astronom\'ia, Facultad de Ingenier\'ia, Ciencia y Tecnolog\'ia, Universidad Bernardo O’Higgins, Av. Viel 1497, Santiago, 8370993, Chile
   \and Specola Vaticana, Vatican Observatory, V00120 Vatican City State, Italy
   \and Universidade de S\~ao Paulo, IAG, Rua do Mat\~ao 1226, Cidade Universit\'aria, S\~ao Paulo 05508-900, Brazil
   \and Instituto de Astronom\'ia Te\'orica y Experimental (CONICET--UNC), Laprida 854, X5000BGR C\'ordoba, Argentina
   \and Observatorio Astron\'omico, Universidad Nacional de C\'ordoba, Laprida 854, X5000BGR C\'ordoba, Argentina.}

   \date{Received 2 April 2026 / Accepted 2 August 2026}

\abstract
{Understanding how the Milky Way disc assembled and evolved requires tracing the coupled evolution of stellar chemistry and orbital structure over time. Open clusters, as coeval stellar populations with well-constrained ages, distances, and chemical properties, provide powerful benchmarks for this purpose.
}
{We aim to characterise the age-dependent chemo-dynamical structure of the Galactic disc and to investigate how open clusters populate dynamical phase space as a function of age.
}
{We analysed a sample of Galactic open clusters using homogeneous chemical abundances from the \emph{Gaia}--ESO Survey together with \emph{Gaia} DR3 phase-space information. Combining cluster ages with orbital parameters and actions, we examined how clusters populate dynamical phase space over the last $\sim4\,\mathrm{Gyr}$.
}
{We find that clusters with similar chemical properties occupy coherent regions of action space, while their orbital structure shows a clear dependence on age. Radial excursions and orbital eccentricities broaden systematically toward older ages, indicating increasingly diverse orbital configurations in the older cluster population. The dispersion in the $[\mathrm{Fe/H}]$--$R_g$ relation also increases with age, consistent with a combination of secular redistribution, survival bias, and dynamically distinct outer-disc populations. In addition, older clusters reach larger vertical amplitudes and vertical actions, revealing a coupled but anisotropic evolution between in-plane and vertical orbital structure.
}
{Open clusters provide precise, age-resolved tracers of the chemo-dynamical structure of the Galactic disc. Our results reveal systematic age-dependent variations in the orbital and chemical properties of the cluster population, consistent with the combined effects of secular evolution, orbital redistribution, and environmentally dependent cluster survival over the last few Gyr.
}

   \maketitle
\nolinenumbers

\section{Introduction}
The Galactic disc preserves a fossil record of its formation and subsequent dynamical evolution in the coupled distributions of stellar chemistry, age, and kinematics. Over the past two decades, large spectroscopic surveys such as RAdial Velocity Experiment (RAVE) \citep{steinmetz2006radial}, Sloan Extension for Galactic Understanding and Exploration (SEGUE) \citep{yanny2009segue}, Apache Point Observatory Galactic Evolution Experiment (APOGEE) \citep{majewski2017apache}, Galactic Archaeology with HERMES (GALAH) \citep{de2015galah}, and Large Sky Area Multi-Object Fiber Spectroscopic Telescope (LAMOST) \citep{cui2012large} together with astrometric and photometric surveys including \emph{Gaia} \citep{brown2016gaia}, The Panoramic Survey Telescope and Rapid Response
System (Pan-STARRS) \citep{schlafly2012photometric}, The Two Micron All Sky Survey (2MASS) \citep{skrutskie2006two}, and VISTA Variables in the Via Lactea and its extension (VVV/VVVX) \citep{minniti2010vista,saito2024vista}, have transformed this field, revealing that the disc is not a simple, axisymmetric system evolving in isolation. Instead, it is a dynamically active structure shaped by spiral arms, the Galactic bar, satellite interactions, and secular heating. These processes redistribute stars in angular momentum and energy, imprinting signatures of radial migration and disc heating that are now observable in chemo-dynamical space \citep{sellwood2002radial,schonrich2009chemical,minchev2013chemodynamical,gomez2013vertical}. 

Field-star studies, particularly with APOGEE and \emph{Gaia}, have demonstrated that stellar populations with similar chemistry can occupy distinct regions in action space, revealing coherent orbital families and age-dependent dynamical structure \citep[e.g][]{mackereth2019dynamical,trick2019galactic,frankel2020keeping,ting2019vertical}. These works show that the disc’s present-day chemo-kinematic architecture encodes both its star-formation history and its dynamical evolution. However, field stars suffer from two fundamental limitations: ages are uncertain for the majority of stars, and individual birth radii are not directly known. As a result, disentangling the relative roles of in-situ chemical evolution, radial migration, and dynamical heating remains challenging.
Open clusters, on the other hand, offer a uniquely powerful, yet still under-exploited, alternative. As coeval stellar populations with well-determined ages and homogeneous chemical abundances, open clusters provide discrete, time-stamped tracers of the disc’s evolution. Their mean phase-space coordinates can be determined with high precision using \emph{Gaia}, allowing robust orbital integration and the derivation of actions and guiding radii. In recent years, homogeneous spectroscopic surveys such as \emph{Gaia}-ESO have delivered multi-element abundances for large samples of open clusters across a wide range of Galactocentric radii and ages \citep{magrini2017gaia,boucher2026gaia}. These data have already been used to map radial abundance gradients and their time evolution \citep[][]{magrini2023gaia,baratella2021gaia}, establishing open clusters as key probes of Galactic chemical evolution.
 
Yet, many cluster-based studies have remained largely one-dimensional, focusing primarily on radial metallicity gradients or age–metallicity relations \citep[e.g.][]{friel1995old,carraro1998galactic,netopil2016metallicity,spina2022mapping,joshi2024study}. The dynamical dimension, that is how clusters move through the disc, how far they have migrated from their birth radii, and whether chemically similar clusters occupy coherent orbital families, 
remains comparatively less explored than field-star work, especially with homogeneous high-resolution chemistry \citep[e.g][]{cantat2020painting,tarricq20213d}. As a result, open clusters have not yet been fully integrated into the chemo-dynamical framework that now dominates field-star studies.

In this paper, we present a systematic cluster-based view of disc chemo-dynamics using the \emph{Gaia}--ESO open cluster programme. By working with age-tagged, chemically homogeneous populations and combining them with orbital actions and guiding radii, we provide an empirical benchmark for chemo-dynamical disc evolution that complements field-star studies. Our central question is: do open clusters organise into coherent structures in joint chemistry--age--dynamics space, and do their ensemble trends reveal age-dependent signatures of secular heating and radial mixing? To address this, we construct cluster centroids in six-dimensional phase space, integrate orbits in a Milky Way potential, and derive orbital parameters and actions $(J_R, J_\phi, J_Z)$ together with guiding radii $R_g$, eccentricities $e$, and vertical amplitudes $Z_{\max}$. We focus on establishing robust empirical correlations and population-level structure, while interpreting these trends in the context of secular evolution and mixing. 
These dynamical quantities are then combined with precise cluster ages and homogeneous multi-element abundances to explore the disc in joint chemical–dynamical–temporal space. By working with open clusters rather than field stars, we gain three decisive advantages: (i) precise ages anchored in stellar evolution \citep{cantat2020painting}, (ii) chemically homogeneous populations that trace discrete star-formation events \citep{bovy2016chemical}, and (iii) well-defined mean orbits that suppress small-scale velocity noise \citep{tarricq20213d}. This allows us to test, in a direct and time-resolved way, whether chemically similar populations trace coherent orbital families, and whether the strength of radial migration and disc heating evolves with age. 

This paper is structured as follows. In Section~\ref{data}, we describe the data used in this study and the quality cuts adopted to select high-quality, chemically homogeneous cluster members. Section~\ref{method} presents the methodology, including the orbit integration procedure and validation using two independent orbital frameworks. In Section~\ref{global_proper}, we present the global orbital and chemo-dynamical properties of the \emph{Gaia}--ESO open-cluster sample analysed in this work. Finally, Sections~\ref{discusion} and~\ref{summary} discuss the results and summarise the main conclusions.

\section{Data}\label{data}
\subsection{Gaia--ESO open-cluster sample}
\begin{figure}
 \includegraphics[width=\columnwidth]{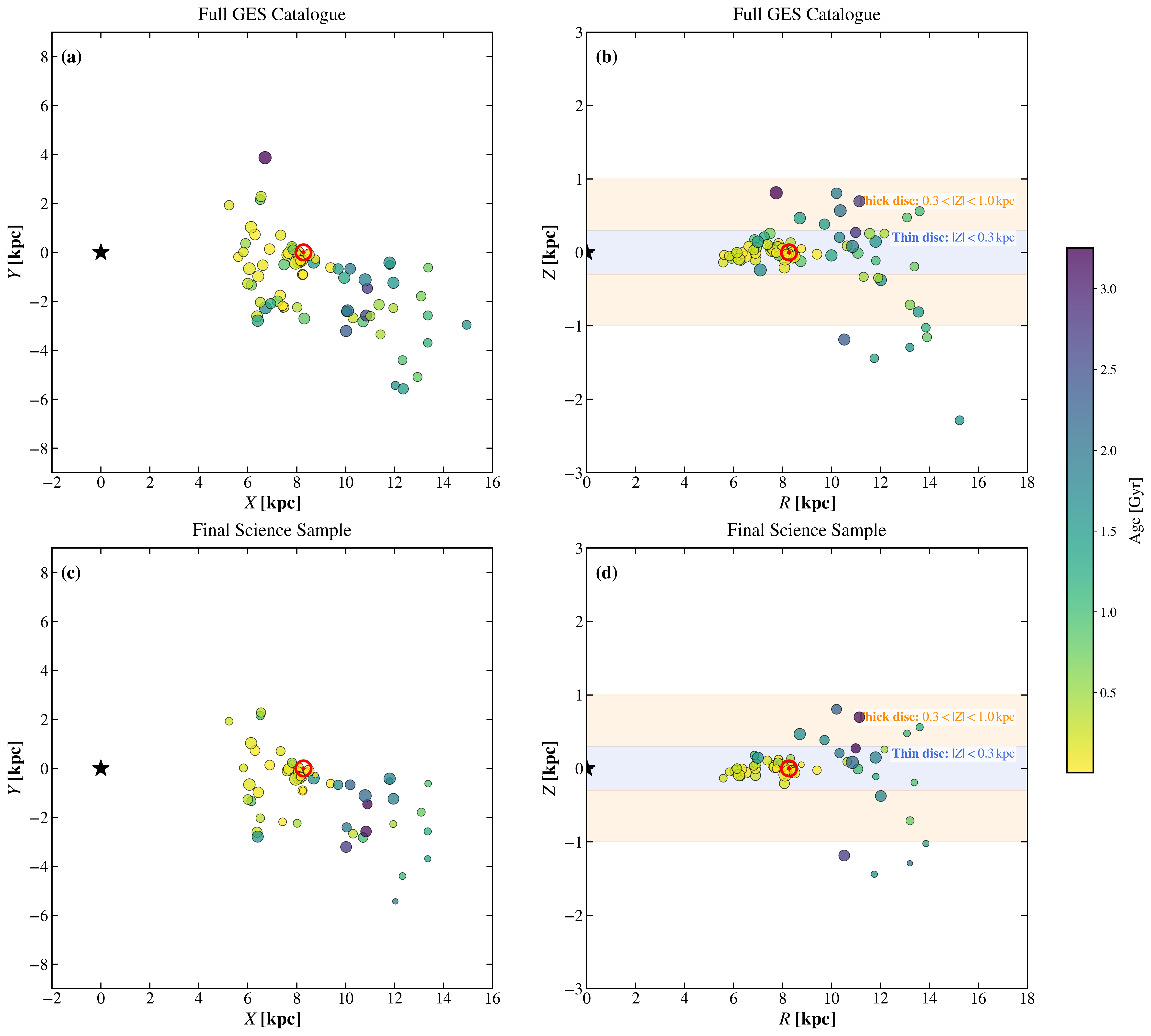}
 \caption{Spatial distribution of the Gaia-ESO open cluster sample in Galactic Cartesian ($X, Y$) and Galactocentric cylindrical ($R, Z$) coordinates. Panels (a) and (b) show the full GES catalogue, while (c) and (d) show the final ALL sample ($N=59$) used in this study. Points are colour-coded by age [Gyr] and scaled by the number of member stars ($\log_{10} N$). The Sun ($\odot$) and Galactic Center (black star) are indicated. The Cartesian reference frame is defined such that the Galactic Centre is located at $(X,Y)=(0,0)$, the Sun lies along negative $X$, and positive $Y$ follows the direction of Galactic rotation. Shaded regions in the right panels represent the nominal thin disc ($|Z| < 0.3$~kpc) and thick disc ($0.3 < |Z| < 1.0$~kpc) layers.}
 \label{fig:1}
\end{figure}

The \emph{Gaia}-ESO Survey (GES) \citep{gilmore2012gaia} is a large public spectroscopic survey carried out with the ESO Very Large Telescope (VLT), designed to provide homogeneous, high-quality stellar parameters and chemical abundances across all major components of the Milky Way, including a dedicated open cluster programme \citep{bragaglia2022gaia,randich2022gaia}. The survey strategy, instrumentation, and data products are described in detail by \citet{gilmore2012gaia} and \citet{randich2013gaia}, with subsequent data releases presented in a series of \emph{Gaia}–ESO Survey papers.
For this work, we make use of the publicly available \emph{Gaia}-ESO catalogue retrieved from the ESO archive\footnote{\url{https://archive.eso.org/scienceportal/home}}, released on 2nd July 2023. The catalogue contains 114,916 unique stars with well-measured spectroscopic parameters. These stars span 218 distinct astrophysical systems, including open clusters, globular clusters, Milky Way field populations, bulge stars, and Magellanic Cloud objects, etc.
The catalogue includes several quality and classification flags that are essential for constructing clean samples tailored to specific scientific goals. In this study, we focus on open cluster members and select targets based on the \texttt{GES\_TYPE} classification. %We include all sources flagged with the extensions \texttt{\_CL} and \texttt{\_OC}. 
We include sources associated with Gaia--ESO open-cluster fields using the \texttt{GES\_TYPE} classification. The \texttt{GES\_TYPE} keyword is a survey provenance descriptor rather than a physical classification of stars or clusters. According to the Gaia--ESO documentation\footnote{\url{https://www.eso.org/rm/api/v1/public/releaseDescriptions/191}}, \texttt{\_CL} identifies stars observed in open-cluster survey fields, while \texttt{\_SD\_OC} denotes standard-field observations of calibrating open clusters. These labels are therefore used only to identify stars associated with cluster-related Gaia--ESO observations; the final cluster sample is subsequently defined through the membership and quality cuts described below.
Sources with \texttt{GES\_TYPE = \_CL} comprise 42,131 spectra corresponding to 81 clusters, while those flagged as \texttt{GES\_TYPE = \_SD\_OC} consist of 1,336 spectra belonging to 10 clusters. The extension reflects how the targets were defined and organised within the survey’s observing strategy. Six clusters are common to both classifications; in these cases, we combined member stars from both flags, retaining only those that satisfy our subsequent quality and membership selection criteria.
The spatial distribution of the clusters used in this analysis is shown in Figure~\ref{fig:1}. The upper panels present the combined open cluster sample selected from \texttt{GES\_TYPE = \_CL} and \texttt{GES\_TYPE = \_SD\_OC}. The left-hand panel shows the cluster distribution in the Galactic Cartesian X--Y plane, while the right-hand panel displays the corresponding X--Z projection. In both panels, the positions of the Galactic Center and the Sun are indicated for reference. In the X--Z plane, we additionally mark the characteristic scale heights of the thin and thick Galactic discs, adopted from the literature, to provide context for the vertical distribution of the clusters \citep{bragaglia2022gaia}. The lower panels show the same projections as the upper panels, but restricted to the final cluster sample that satisfies all quality and membership selection criteria described in Section~\ref{quality_cut}. In both the upper and lower panels, clusters are colour-coded by age, with the accompanying colour bar indicating the age distribution of the sample. This representation highlights both the radial and vertical coverage of the \emph{Gaia}--ESO open cluster sample, as well as the changes in spatial and age distributions introduced by the final selection.
\subsection{Quality cuts and sample definition}
\label{quality_cut}

To construct a homogeneous open cluster sample suitable for chemo-dynamical analysis, we applied a two-stage quality control procedure to the \emph{Gaia}--ESO Survey data. This process consists of 
stellar-level filtering followed by cluster-level aggregation, defining three nested samples: \textsc{all} ($N=59$), \textsc{science} ($N=50$), and \textsc{gold} ($N=23$).

stellar-level selection: At the stellar level, we retained only high-confidence cluster members. We retained only stars with Gaia--ESO Survey three-dimensional membership probabilities of $\mathrm{MEM3D} \geq 0.9$, thereby minimising contamination from field stars. Spectra were required to have signal-to-noise ratios $\mathrm{SNR} \geq 20$. We further required finite and well-converged atmospheric parameters, as indicated by the neural-network quality flags $\mathrm{NN\_TEFF} \geq 2$, $\mathrm{NN\_LOGG} \geq 2$, and $\mathrm{NN\_FEH} \geq 1$. Only stars with measured radial velocities and uncertainties $\sigma_{\mathrm{RV}} \leq 5~\mathrm{km\,s^{-1}}$ were retained. Where iron abundances were available, we required uncertainties $\sigma_{\mathrm{[Fe/H]}} \leq 0.20~\mathrm{dex}$.

Cluster-level aggregation: Stellar measurements passing the stellar-level selection were combined to derive mean cluster properties. We adopted robust statistical estimators, using the median to define the central cluster properties and the scaled median absolute deviation ($1.4826 \times \mathrm{MAD}$) as a dispersion estimator, to minimize sensitivity to outliers. The scaling factor converts the MAD into a consistent estimator of the standard deviation under the assumption of approximately Gaussian errors. 
For each cluster, we derived robust central properties using median-based estimators, including the sky position, radial velocity, velocity dispersion, and chemical abundances for all available species.

Sample definitions: The criteria for each nested sample are summarized in Table~\ref{tab:sample_cuts}. The \textsc{all} sample includes all clusters with at least one star passing the stellar-level quality cuts. The \textsc{science} sample, used for our main population-level analysis, adopts relaxed but physically motivated criteria ($N_{\star} \geq 2$) to ensure broad coverage in age and Galactocentric radius. Finally, the \textsc{gold} sample provides a high-fidelity subset for benchmarking, requiring higher membership ($N_{\star} \geq 3$ or $4$) and tight kinematic constraints ($\sigma_{\mathrm{RV}} \leq 3~\mathrm{km\,s^{-1}}$) consistent with the virial states of bound clusters. The final properties for the full sample are detailed in Table~\ref{quality_cut_fig} in the appendix.

\begin{table}
\centering
\caption{Summary of selection criteria for the nested cluster samples.}
\label{tab:sample_cuts}
\begin{tabular}{lccc}
\hline
\hline
Criterion & \textsc{all} & \textsc{science} & \textsc{gold} \\
\hline
Min. stars ($N_{\mathrm{Fe}}$)  & $\geq 1$ & $\geq 2$   & $\geq 3$ \\
Fe dispersion ($\sigma_{\mathrm{[Fe/H]}}$) & --- & $\leq 0.18$ dex & $\leq 0.18$ dex \\
Min. stars ($N_{\mathrm{RV}}$)  & $\geq 1$ & $\geq 2$   & $\geq 4$ \\
Vel. dispersion ($\sigma_{\mathrm{RV}}$) & --- & $\leq 8.5~\mathrm{km\,s^{-1}}$ & $\leq 3.0~\mathrm{km\,s^{-1}}$ \\
$\alpha$-element requirement & No & No & $\geq 2$ species \\
\hline
\textbf{Final count ($N_{\mathrm{clus}}$)} & \textbf{59} & \textbf{50} & \textbf{23} \\
\hline
\end{tabular}
\end{table}

\section{Method}\label{method}

In this section, we describe the methodology adopted to analyse the chemo-dynamical properties of the open cluster sample. We outline the construction of the six-dimensional (6D) phase-space coordinates for each cluster and describe how the derived orbital parameters were validated using two independent orbit-integration frameworks, \textsc{galpy} \citep{bovy2015galpy} and \texttt{GravPot16}\footnote{https://gravpot.utinam.cnrs.fr} \citep[see][for a detailed discussion of the code]{trincado2017structure,fernandez2020dynamical}.

\subsection{\emph{Gaia} cross-matching and 6D phase-space construction}\label{6D_phase}

To link the chemical properties of open clusters to their Galactic dynamics, we cross-matched our cluster sample with \emph{Gaia} DR3 and recent homogeneous open cluster catalogues \citep{hunt2023improving}. The cross-matching was performed using cluster centroids on the sky, adopting angular radii chosen to account for the extended spatial morphologies of open clusters and small coordinate differences between catalogues reported in the literature.

We adopted the mean \emph{Gaia} DR3 cluster parallaxes and proper motions compiled by \citet{hunt2023improving}. These astrometric parameters were combined with \emph{Gaia}--ESO Survey radial velocities and spectroscopically determined cluster centroids to construct full 6D phase-space coordinates, $(\alpha, \delta, d, \mu_{\alpha}^{\ast}, \mu_{\delta}, v_{\rm los})$, for each cluster. We adopted \emph{Gaia}--ESO radial velocities in order to maintain internal consistency with the spectroscopic membership determination and chemical-abundance analysis, since the velocities are derived from the same spectra used throughout this work. We do not imply that \emph{Gaia}--ESO radial velocities are intrinsically more reliable than \emph{Gaia} RVS measurements; rather, the \emph{Gaia}--ESO dataset provides a homogeneous spectroscopic framework for the analysed cluster sample.

Cluster ages were adopted from the homogeneous catalogue of \citet{hunt2023improving}. Using a single age source ensures internal consistency across the sample and avoids heterogeneous systematics associated with combining values from multiple literature studies. For very young clusters, conversion of logarithmic age uncertainties into linear ages in Gyr can produce apparently negligible errors after truncation in tabulated values, although the underlying uncertainties remain finite.
The resulting six-dimensional phase-space information and homogeneous age estimates form the basis for the orbit integrations and action calculations described in the following section. We emphasize, however, that the \emph{Gaia}--ESO open-cluster sample is not a complete census of the Galactic open-cluster population. The survey selection function, optical targeting strategy, and extinction-dependent incompleteness may affect the spatial and age distributions of the analysed clusters. The trends presented in this work should therefore be interpreted as empirical relations within the \emph{Gaia}--ESO cluster sample rather than as volume-complete measurements of the Milky Way open-cluster population.

\begin{table*}[!t]
\centering
\caption{Derived orbital parameters for the SCIENCE sample of Gaia--ESO
open clusters. Only the first five rows are shown; the full
machine-readable table is available electronically.}
\label{tab:orbital_parameters_science}

\tiny
\setlength{\tabcolsep}{1pt}
\renewcommand{\arraystretch}{0.70}

\resizebox{\textwidth}{!}{%
\begin{tabular}{lcccccccccccccccccccccc}
\toprule
GES\_FLD &
$d$ &
$\sigma_d$ &
$\log\,\mathrm{Age}$ &
$\sigma_{\log\mathrm{Age}}$ &
$\mathrm{Age}$ &
$\sigma_{\mathrm{Age}}$ &
$R_{\mathrm{peri}}$ &
$\sigma_{R_{\mathrm{peri}}}$ &
$R_{\mathrm{apo}}$ &
$\sigma_{R_{\mathrm{apo}}}$ &
$e$ &
$\sigma_e$ &
$Z_{\max}$ &
$\sigma_{Z_{\max}}$ &
$J_R$ &
$\sigma_{J_R}$ &
$J_{\phi}$ &
$\sigma_{J_{\phi}}$ &
$J_Z$ &
$\sigma_{J_Z}$ &
$R_g$ &
$\sigma_{R_g}$ \\
\midrule
Blanco1 & 234.390 & 0.094 & 8.239 & 0.309 & 0.173 & 0.140 &
8.079 & 0.004 & 8.713 & 0.001 & 0.037 & 0.000 & 0.215 & 0.001 &
1.867 & 0.021 & 1938.128 & 0.520 & 1.551 & 0.019 & 8.371 & 0.002 \\
Br21 & 5294.645 & 152.808 & 8.945 & 0.255 & 0.881 & 0.606 &
12.240 & 0.135 & 14.211 & 0.215 & 0.074 & 0.010 & 0.235 & 0.007 &
10.722 & 2.986 & 2888.963 & 21.462 & 0.949 & 0.041 & 13.157 & 0.111 \\
Br22 & 5334.363 & 128.064 & 8.897 & 0.263 & 0.789 & 0.545 &
10.910 & 0.257 & 14.559 & 0.221 & 0.143 & 0.017 & 0.789 & 0.020 &
38.060 & 9.047 & 2749.343 & 25.489 & 7.940 & 0.281 & 12.436 & 0.131 \\
Br30 & 4462.350 & 73.578 & 8.663 & 0.180 & 0.460 & 0.190 &
11.439 & 0.114 & 12.697 & 0.469 & 0.050 & 0.019 & 0.278 & 0.007 &
4.426 & 4.838 & 2672.182 & 40.891 & 1.518 & 0.030 & 12.039 & 0.209 \\
Br31 & 5864.814 & 174.234 & 9.075 & 0.228 & 1.190 & 0.708 &
10.954 & 0.180 & 13.708 & 0.174 & 0.111 & 0.007 & 0.587 & 0.015 &
22.195 & 3.176 & 2693.462 & 29.961 & 5.216 & 0.146 & 12.148 & 0.153 \\
\bottomrule
\end{tabular}%
}

\tablefoot{
Columns give heliocentric distance ($d$; pc), age (Gyr), pericentre and
apocentre radii (kpc), orbital eccentricity, maximum vertical excursion
(kpc), orbital actions ($J_R$, $J_{\phi}$, $J_Z$;
kpc\,km\,s$^{-1}$), and guiding radius (kpc). Uncertainties are
1$\sigma$. Only the first five rows are shown; the full
machine-readable table is available electronically.
}

\end{table*}

%////////////////////////////////////////////////////////////////////////////////////////////////////

\subsection{Galactic potential and orbit integration}
\label{sec:orbits}

To characterise the Galactic orbits of the open clusters in our sample, we integrated their trajectories in a fixed, axisymmetric Milky Way potential using the \textsc{galpy} package \citep{bovy2015galpy}. 
We adopted the \texttt{MWPotential2014} model, which consists of a three-component mass distribution including a bulge, disc, and dark matter halo, and has been widely used in Galactic archaeology studies of disc stellar populations and open clusters.
 
We assumed a Galactocentric distance of the Sun $R_{0} = 8.2~\mathrm{kpc}$ \citep{Gravity2019}, a circular velocity at the solar radius $V_{0} = 232~\mathrm{km\,s^{-1}}$ \citep{mcmillan2016mass}, and a solar height above the mid-plane $z_{0} = 25~\mathrm{pc}$ \citep{bennett2019vertical}. The Solar peculiar motion with respect to the local standard of rest was adopted as $(U_\odot, V_\odot, W_\odot) = (11.1, 12.24, 7.25)\,\mathrm{km\,s^{-1}}$ 
following \citep{schonrich2010local}. These values are consistent with recent observational constraints and are commonly adopted in \emph{Gaia}-era dynamical analyses.
Orbits were initialised using the full six-dimensional phase-space information of each cluster centroid, expressed in equatorial coordinates and heliocentric observables $(\alpha, \delta, d, \mu_{\alpha}^{\ast}, \mu_{\delta}, v_{\rm los})$,
where distances, proper motions, and radial velocities were taken from the \emph{Gaia} DR3 cross-matched catalogue of \citet{hunt2023improving} and the \emph{Gaia}--ESO Survey, as described in Section~\ref{6D_phase}. The transformation to a Galactocentric reference frame was handled internally by \textsc{galpy}.

Each orbit was integrated forward in time for $0.5~\mathrm{Gyr}$ with a temporal resolution of $1~\mathrm{Myr}$. This integration interval should be regarded as a first-order numerical sampling of the present-day orbit in the adopted Galactic potential, sufficient to estimate pericentre ($R_{\rm peri}$), apocentre ($R_{\rm apo}$), $e$, and maximum vertical excursion from the Galactic plane ($Z_{\max}$) for the purposes of this work. We emphasise that the goal is not to reconstruct the full dynamical history of each cluster over its lifetime but rather to derive consistent present-day orbital diagnostics for comparative chemo-dynamical analysis. From the integrated orbits, we derived classical orbital parameters, including $R_{\rm peri}$,  $R_{\rm apo}$, orbital eccentricity $e$, and $Z_{\max}$.

In addition to these quantities, we computed orbital actions $(J_{R}, J_{\phi}, J_{Z})$ using the St\"ackel approximation as implemented in \textsc{galpy} \citep{BinneyTremaine2008}. 
A focal length parameter of $\delta = 0.45$ was adopted, which is appropriate for disc-like orbits in the adopted potential. The guiding-centre radius $R_{\rm g}$ was also calculated for each cluster and is used throughout this work as a proxy for angular-momentum-defined orbital radius.

Uncertainties on all derived orbital parameters were estimated via Monte Carlo error propagation. For each cluster, we generated 500 realisations of the input phase-space coordinates by sampling the observed distances, proper motions, and radial velocities assuming Gaussian uncertainties. Orbits were re-integrated for each realisation, and the final orbital parameters were taken as the median of the resulting distributions, with uncertainties estimated from their standard deviations. This approach ensures that observational uncertainties are consistently propagated through the non-linear orbit integration and action computation. We note that the Monte Carlo sampling treats input uncertainties as independent Gaussians; for cluster centroids, the impact of astrometric covariances is expected to be sub-dominant relative to the intrinsic cluster-to-cluster scatter in the trends discussed here, but this approximation may modestly affect uncertainties for individual clusters.
Table~\ref{tab:orbital_parameters_science} summarizes the full set of orbital parameters derived in this work, together with their associated uncertainties. We note that these calculations are performed in a static, axisymmetric Galactic potential and therefore do not account for secular evolution or non-axisymmetric perturbations such as the bar and spiral arms. The resulting orbital parameters should thus be interpreted as present-day diagnostics in the adopted potential, particularly for the youngest clusters in the sample, rather than as detailed reconstructions of their full dynamical histories.

\subsection{Validation with an independent orbit-integration framework}
\label{sec:validation}

\begin{figure}
 \includegraphics[width=\columnwidth]{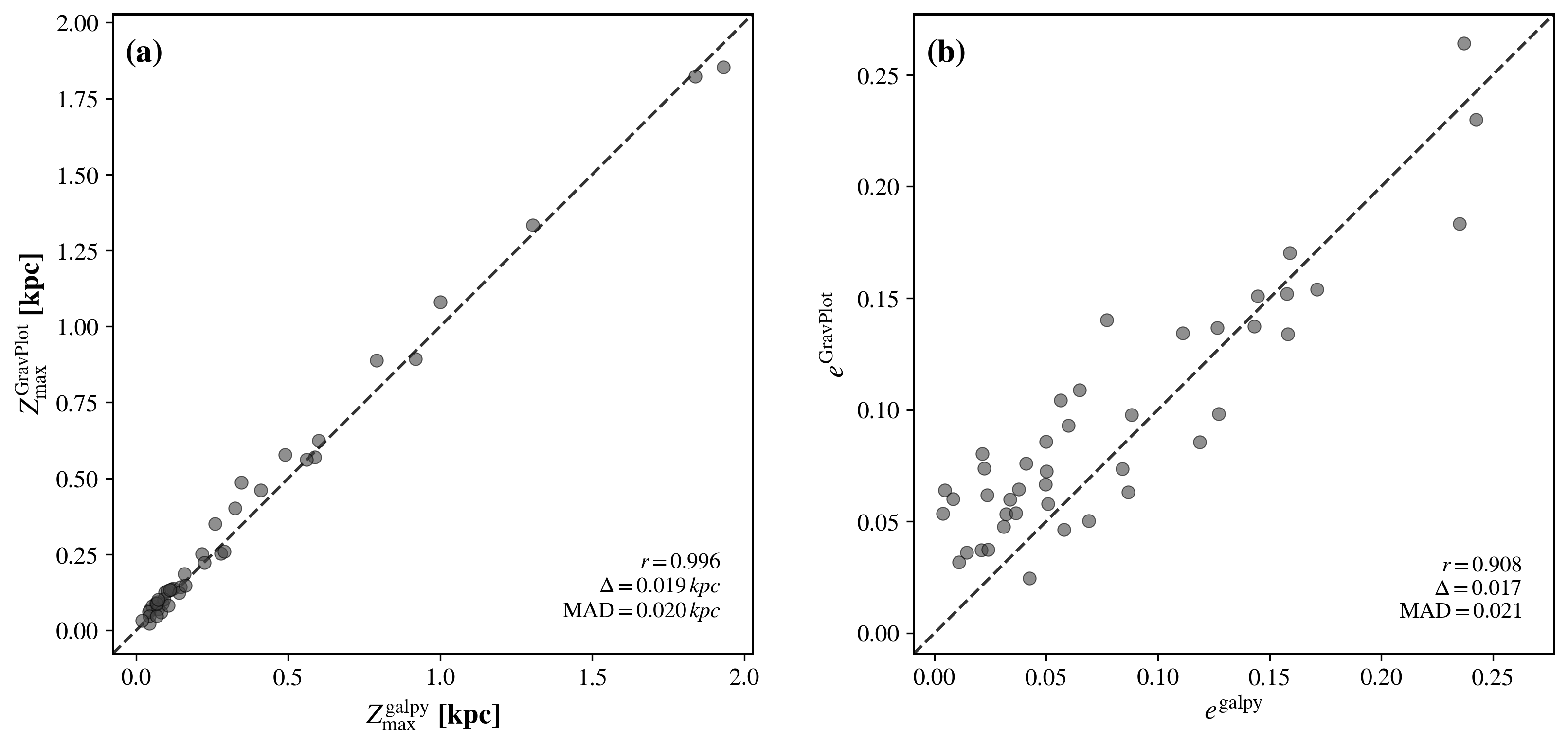}
 \caption{ Comparison of orbital parameters derived from \texttt{galpy} and \texttt{GravPlot}. Panel (a) compares the maximum vertical amplitude ($Z_{\max}$), and panel (b) compares the orbital eccentricity ($e$). The dashed line in each panel indicates the 1:1 relation. The two orbit-integration frameworks show excellent agreement, with Pearson correlation coefficients of $r = 0.996$ for $Z_{\max}$ and $r = 0.908$ for $e$. The median offset ($\Delta$) and the Median Absolute Deviation (MAD) are also reported in each panel. }
 
 \label{fig:2}
\end{figure}

Since the Galactic potential is not known exactly, and the inferred orbital properties may depend to some degree on the adopted potential, we also computed orbits with \texttt{GravPot16} \citep{trincado2017structure,fernandez2020dynamical}, which includes non-axisymmetric perturbations associated with a rotating bar in the inner Galaxy. The \texttt{GravPot16} integrations were carried out independently by José Fernández-Trincado as an external validation of the derived orbital properties, providing a fully independent cross-check of our results. For each cluster, the resulting orbital parameters were compared directly to those obtained with \textsc{galpy} \citep{bovy2015galpy}. We find excellent agreement between the two frameworks for all key orbital quantities considered in this work, including guiding radii, pericentre and apocentre distances, orbital eccentricities, and maximum vertical excursions, with median offsets consistent with zero. As representative examples, the maximum vertical excursion $Z_{\max}$ and the orbital eccentricity exhibit Pearson correlation coefficients of $r = 0.996$ and $r = 0.908$, respectively. Figure~\ref{fig:2} illustrates these comparisons, demonstrating that the orbital parameters derived with \textsc{galpy} are robust against the choice of orbit-integration framework. This agreement confirms that our results are not driven by code-specific implementations and provides confidence that the inferred chemo-dynamical trends discussed in the following sections reflect genuine properties of the open cluster population rather than methodological artefacts.

\subsection{Guiding radii and radial-mixing diagnostics}

To characterise the angular-momentum distribution of the open cluster sample, we computed the guiding radius, $R_g$, for each orbit. In an axisymmetric Galactic potential, $R_g$ corresponds to the radius of a circular orbit with the same angular momentum $L_z$ and provides a phase-independent proxy for the characteristic orbital radius. Guiding radii were derived self-consistently within \textsc{galpy} \citep{bovy2015galpy}.

Throughout this work, we use $R_g$ as an angular-momentum-based coordinate to examine how cluster chemistry and age map onto orbital families. We further quantify radial excursions using the instantaneous offset from the guiding radius, $R - R_g$. In a static axisymmetric potential this quantity primarily traces epicyclic phase and the amplitude of radial excursions (radial blurring), rather than changes in angular momentum. We therefore adopt a conservative interpretation: trends involving $|R - R_g|$ are used as empirical diagnostics of in-plane orbital excursion and mixing, without attempting to isolate churning (i.e. net angular-momentum change).
Uncertainties on all derived orbital parameters and actions were estimated via Monte Carlo sampling as described above. All quoted uncertainties reflect this full error propagation unless stated otherwise.

\section{Global orbital and chemo-dynamical properties of the \emph{Gaia}--ESO open cluster sample}\label{global_proper}
In this section, we present the global orbital and chemo-dynamical properties of the \emph{Gaia}–ESO open cluster sample analysed in this study. Using homogeneous cluster ages, multi-element abundances, and orbital parameters derived from \emph{Gaia} DR3 astrometry and \emph{Gaia}–ESO spectroscopy, we characterise the distribution of open clusters in dynamical, chemical, and age–orbital parameter spaces. We describe the observed trends in guiding radius, orbital eccentricity, vertical excursion, and orbital actions, and examine their connections with cluster age and chemical composition. Throughout this section, we focus on empirically establishing correlations and population-level behaviours, while comparisons with theoretical expectations and previous studies are provided where relevant; a broader physical interpretation is deferred to Section~\ref{discusion}.
\begin{table*}
\centering
\caption{Summary of empirical secular heating and radial-mixing diagnostics for the \textsc{science} sample.}
\label{tab:final_stats}
\begin{tabular}{lllc}
\hline
\hline
Dynamical Process & Metric & Value / Slope & Significance ($p$) \\
\hline
\textit{Radial evolution (in-plane)} & & & \\
Radial excursion amplitude (blurring proxy) & $m(|R-R_g|,\mathrm{Age})$ & $0.195 \pm 0.036~\mathrm{kpc\,Gyr^{-1}}$ & $< 10^{-3}$ \\
Eccentricity--age slope$^{a}$ & $m(e,\mathrm{Age})$ & $0.081 \pm 0.013~\mathrm{Gyr^{-1}}$ & $3.1 \times 10^{-9}$ \\
Robust slope (Theil--Sen) & $m_{\mathrm{TS}}(e,\mathrm{Age})$ & $0.056~\mathrm{Gyr^{-1}}$ & --- \\
Chemo-dynamical coupling & $r_s(J_R, |\Delta \mathrm{[Fe/H]}|)$ & $0.378$ & $6.8 \times 10^{-3}$ \\
Mean metallicity gradient & $\nabla_{R_g}$ & $-0.076 \pm 0.008~\mathrm{dex\,kpc^{-1}}$ & --- \\
\hline
\textit{Vertical evolution (out-of-plane)} & & & \\
Vertical thickening slope & $m(Z_{\max},\mathrm{Age})$ & $0.033 \pm 0.010~\mathrm{kpc\,Gyr^{-1}}$ & $< 10^{-10}$ \\
Age--$Z_{\max}$ correlation & $\rho(\mathrm{Age}, Z_{\max})$ & $0.80$ & $< 10^{-10}$ \\
\hline
\multicolumn{4}{l}{$^{a}$ Empirical slope from ODR accounting for uncertainties in both variables; a robust Theil--Sen estimator yields a consistent slope.}
\end{tabular}
\tablefoot{Slopes are descriptive empirical trends derived from Orthogonal Distance Regression (ODR) accounting for uncertainties in both variables unless otherwise stated. Correlations are reported as Spearman rank coefficients ($\rho$ or $r_s$). The metallicity gradient uncertainty reflects the formal ODR error; intrinsic scatter is discussed separately in the text.}

\end{table*}
\subsection{Global orbital properties}
\begin{figure}
 \includegraphics[width=\columnwidth]{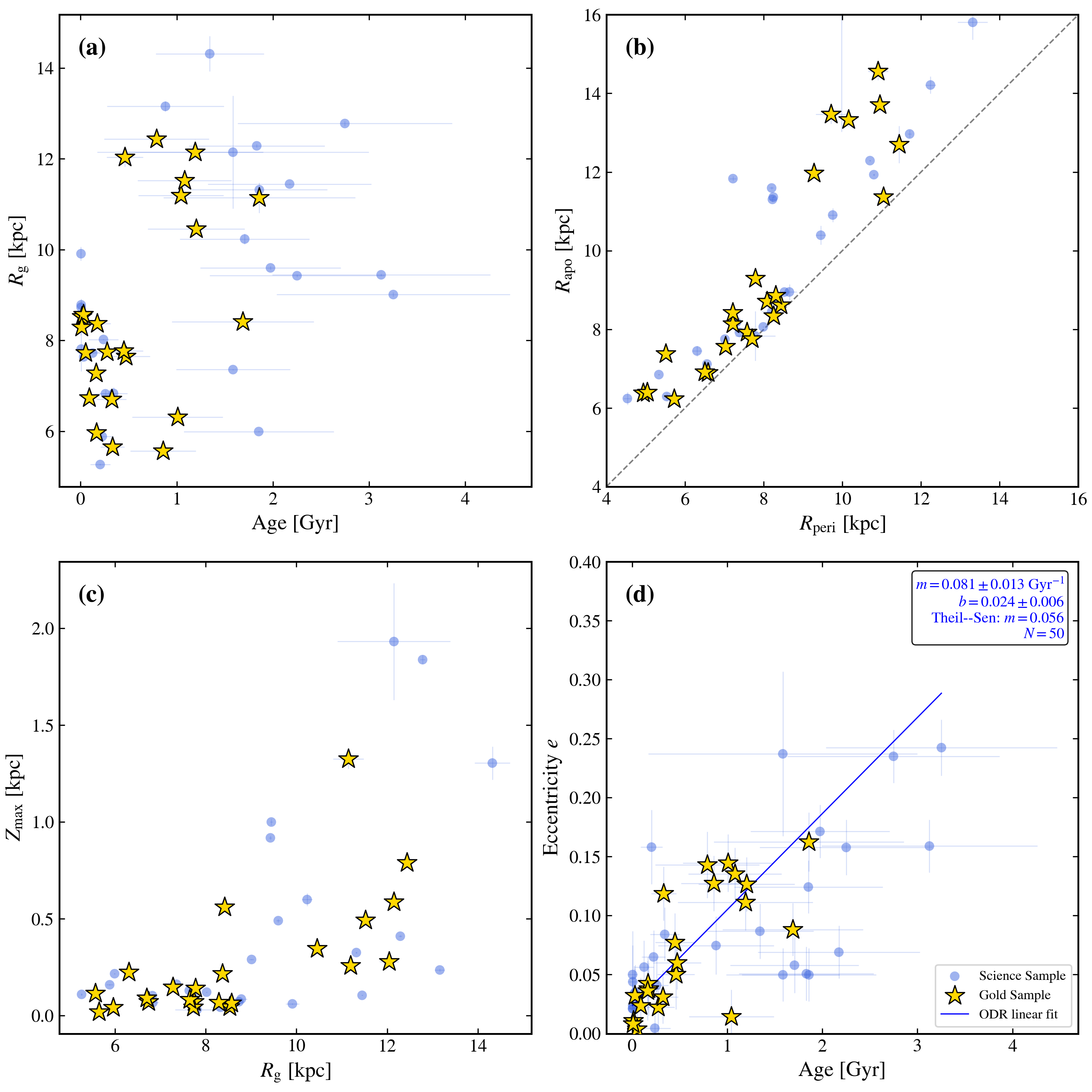}
 \caption{Global orbital properties of the \emph{Gaia}--ESO open cluster sample. 
Panels show (a) guiding radius as a function of age, (b) pericentre versus apocentre radii, (c) maximum vertical excursion as a function of guiding radius, and (d) orbital eccentricity as a function of age. 
The SCIENCE sample is shown with filled circles, while the high-quality GOLD subsample is highlighted with star symbols. The full ALL sample is shown in the background as open grey circles. In panel (d), the solid line represents the best-fitting linear relation between eccentricity and age, obtained using orthogonal-distance regression to account for uncertainties in both variables. 
}

 \label{fig:3}
\end{figure}

We begin by presenting the global orbital properties of the \emph{Gaia}–ESO open cluster sample, providing an overview of the dynamical characteristics spanned by the clusters analysed in this work. Figure~\ref{fig:3} summarises these properties in four complementary projections of orbital parameter space. The full SCIENCE sample is shown with filled circles, while the higher-quality GOLD subsample is highlighted with star symbols. The panels illustrate the distribution of cluster angular momentum, radial excursions, vertical structure, and dynamical heating as a function of age, thereby offering a compact visual summary of the range of orbital families represented in the sample. Together, these diagnostics characterise how open clusters populate the Galactic disc in phase space and establish the empirical basis for the more detailed age–orbital and chemo-dynamical relations explored in the following subsections.
\subsubsection{Guiding radii and angular-momentum redistribution}
The upper-left panel of Figure~\ref{fig:3} shows the guiding radius, $R_{\mathrm{g}}$, as a function of cluster age. 
The sample spans a broad radial range, from $R_{\mathrm{g}} \sim 5$ to $\sim 14$~kpc, indicating that the clusters probe both the inner and outer Galactic disc. 
At fixed age, a substantial dispersion in $R_{\mathrm{g}}$ is observed, particularly for clusters older than $\sim 1$~Gyr. %This dispersion implies that clusters of similar age occupy a range of angular-momentum states. Such behaviour is qualitatively consistent with radial migration driven by angular-momentum exchange, which can redistribute stellar populations in radius without necessarily increasing orbital eccentricity \citep{sellwood2002radial,schonrich2009chemical}.
This dispersion shows that clusters of similar age occupy a broad range of angular-momentum states. However, the observed distribution should not be interpreted uniquely as evidence for radial migration. In particular, the scarcity of older clusters at small $R_{\mathrm{g}}$ may reflect a combination of secular orbital redistribution \citep{sellwood2002radial,schonrich2009chemical}, enhanced tidal disruption in the inner disc, encounters with giant molecular clouds \citep{lamers2005disruption,gieles2006star,kruijssen2011modelling}, and extinction-dependent selection effects toward the inner Galaxy \citep{cantat2020painting}. We therefore interpret the broad range in $R_g$ primarily as an empirical signature of the angular-momentum diversity of the surviving cluster population, rather than as a direct tracer of cluster migration histories.
The GOLD subsample reproduces the same distribution, indicating that the observed spread is not driven by poorly constrained orbital solutions or astrometric uncertainties.

\subsubsection{Radial orbital excursions}
The upper-right panel of Figure~\ref{fig:3} compares pericentre ($R_{\rm peri}$) and apocentre ($R_{\rm apo}$) radii.
Most clusters lie close to the one-to-one relation, indicating modest radial excursions around their guiding radii and predominantly disc-like orbits.
Quantitatively, the typical radial excursion, defined as $\Delta R = R_{\rm apo}-R_{\rm peri}$, is $\sim1.1\,\mathrm{kpc}$ for the combined sample, with a 16th–84th percentile range of $0.32$–$2.91\,\mathrm{kpc}$.
In fractional terms, the median excursion corresponds to $\Delta R/R_g \approx 0.11$, indicating that most clusters remain tightly confined around their guiding radii.
However, a subset of clusters, particularly those at larger Galactocentric radii, exhibit broader radial excursions.
This behaviour is consistent with gradual dynamical heating induced by interactions with transient spiral structure, giant molecular clouds, and the Galactic bar \citep{jenkins1990spiral,minchev2013chemodynamical}.
Importantly, even clusters with the largest excursions remain far from halo-like kinematics, confirming that the sample traces disc populations throughout its age range.

\begin{figure}
 \includegraphics[width=\columnwidth]{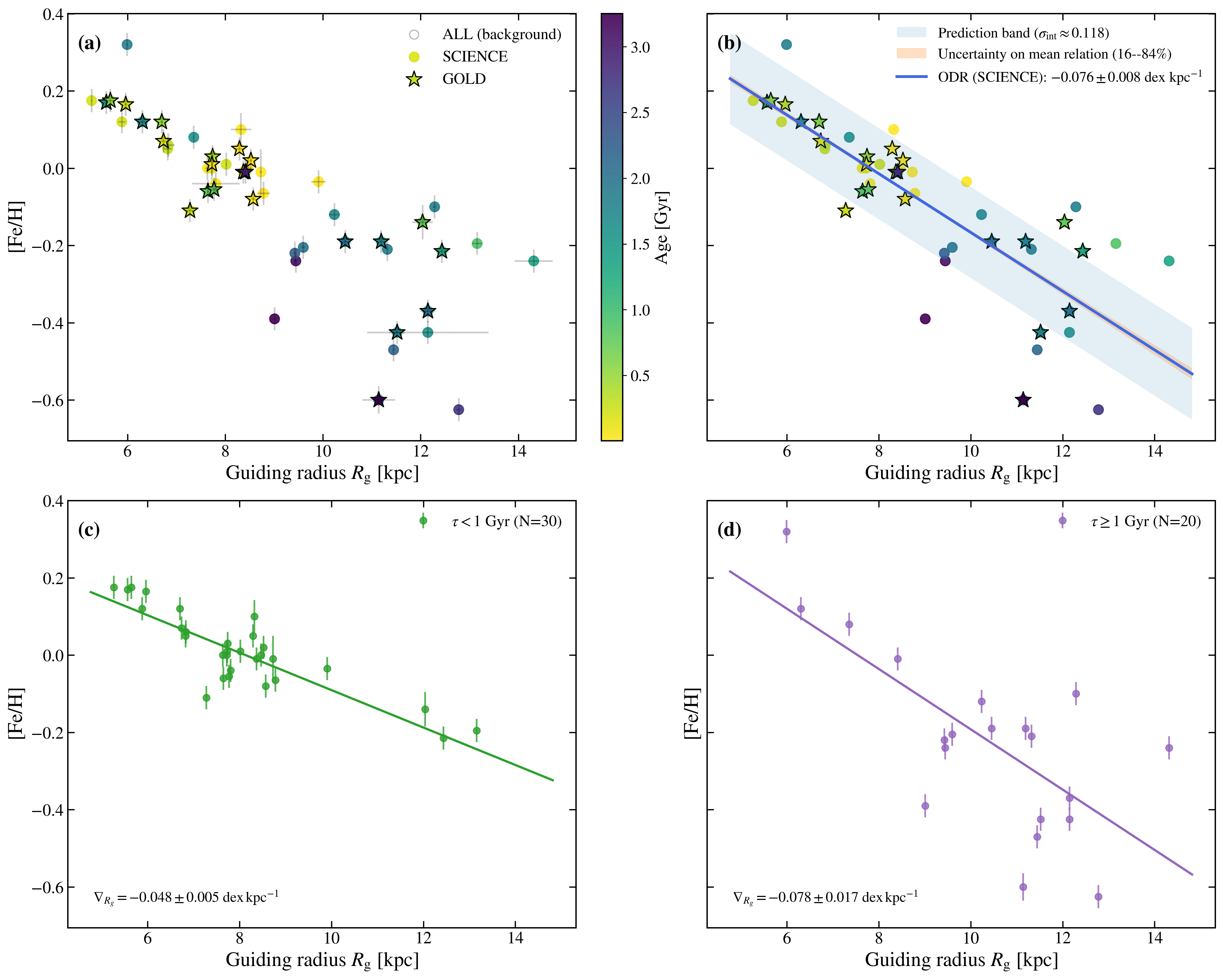}
 \caption{Mean cluster metallicity, $[\mathrm{Fe/H}]$, as a function of guiding radius $R_{\mathrm{g}}$ for the \emph{Gaia}--ESO open cluster sample.
(a) Distribution of clusters in the $[\mathrm{Fe/H}]$--$R_{\mathrm{g}}$ plane, showing the full catalogue (open grey circles), the SCIENCE sample (filled circles), and the high-quality GOLD subsample (star symbols). Points are colour-coded by cluster age. (b) Linear fit to the SCIENCE sample. The narrow shaded region indicates the uncertainty on the mean relation (16--84 percentile), while the broader shaded region represents the prediction band including intrinsic scatter. (c) Same relation restricted to clusters younger than 1~Gyr ($N=30$), with a weighted linear fit. (d) Same as panel~(c), but for clusters older than or equal to 1~Gyr ($N=20$). 
}

 \label{fig:4}
\end{figure}
\subsubsection{Vertical structure of the disc}

The lower-left panel of Figure~\ref{fig:3} shows the maximum vertical excursion, $Z_{\max}$, as a function of guiding radius.
Most clusters are confined within $Z_{\max} \lesssim 0.3$~kpc, consistent with a thin-disc origin.
A small number of systems reach $Z_{\max} \gtrsim 1$~kpc, indicative of dynamically hotter or thicker-disc orbits. 
A mild positive trend is visible, in the sense that clusters at larger $R_{\mathrm{g}}$ tend to reach larger vertical amplitudes. This behaviour is qualitatively consistent with a flaring outer disc \citep{chrobakova2022warp,uppal2024warp}, although the present sample does not allow us to separate flaring from selection effects and age-dependent survival biases. We therefore treat the $Z_{\max}$--$R_g$ relation as descriptive and use the vertical action $J_Z$ as the more robust dynamical diagnostic of vertical heating in the following analysis.

\subsubsection{Orbital eccentricity as a function of age}
The lower-right panel of Figure~\ref{fig:3} shows orbital eccentricity as a function of cluster age. We find a strong monotonic correlation between age and eccentricity (Spearman $\rho = 0.72$, $p = 3.1 \times 10^{-9}$), indicating that older clusters preferentially occupy more eccentric orbits. Fitting a 
linear relation using orthogonal-distance regression, which accounts for uncertainties in both axes and allows for intrinsic scatter, yields an empirical slope of $m = 0.081 \pm 0.013~\mathrm{Gyr^{-1}}$ and intercept $b = 0.024 \pm 0.006$ over the sampled age range (0--3.25~Gyr). A robust Theil--Sen \citep{sen1968estimates} estimator gives a consistent slope of $m \simeq 0.056~\mathrm{Gyr^{-1}}$, confirming the stability of the inferred trend. The youngest clusters (age $<1$~Gyr; $N=30$) are dynamically cold, with a mean eccentricity of $\langle e \rangle = 0.051 \pm 0.007$, where the uncertainty represents the standard error of the mean (SEM). In contrast, clusters older than $2$~Gyr ($N=5$) exhibit systematically higher eccentricities, $\langle e \rangle = 0.173 \pm 0.032$, with the two distributions differing significantly (Mann--Whitney $p = 3.7 \times 10^{-4}$). 
In an axisymmetric potential, this behaviour primarily traces an increase in radial excursion amplitude (dynamical blurring) rather than direct angular-momentum redistribution. 
The broadening of the eccentricity distribution with age is consistent with cumulative in-plane dynamical evolution of the cluster population \citep[e.g.][]{aumer2009kinematics, seabroke2007revisiting}. However, the observed distribution likely reflects a combination of secular heating, preferential survival of dynamically hotter clusters at large radii, and the enhanced destruction of clusters in the inner Galactic disc \citep{lamers2005disruption,gieles2006star,kruijssen2011modelling}. We therefore interpret the eccentricity broadening as an empirical signature of age-dependent orbital evolution rather than as evidence for a single dominant heating mechanism.
A small subset of older clusters retains relatively low eccentricities ($e < 0.1$), indicating that some systems can preserve nearly circular orbits despite their advanced ages. However, the present diagnostics do not uniquely distinguish between secular orbital redistribution, orbital blurring, and selection effects. We therefore avoid interpreting these clusters as direct evidence for radial migration processes operating without significant orbital heating.

\subsection{Metallicity as a function of guiding radius}
\label{sec:feh_rg}

In Figure~\ref{fig:4} we present a summary of the radial metallicity structure of the Galactic disc traced by the \emph{Gaia}--ESO open cluster sample, and show how the interpretation of these data changes once we (i) distinguish uncertainty in the mean metallicity gradient from the intrinsic cluster-to-cluster dispersion, and (ii) examine whether this dispersion depends on age. We use the guiding radius as an angular-momentum-based proxy for the characteristic orbital radius, which is less sensitive to epicyclic phase than the instantaneous Galactocentric radius. While $R_g$ is not a direct measurement of birth radius, it provides a useful coordinate for comparing chemistry with orbital families.

The upper-left panel shows the basic observable: mean cluster metallicity, $[\mathrm{Fe/H}]$, as a function of $R_{\mathrm{g}}$ for the SCIENCE sample (circles) and the high-quality GOLD subsample (stars), and the ALL sample in the background (open circles) with points colour-coded by age. A clear negative trend is immediately apparent, consistent with the long-established radial metallicity gradient traced by open clusters in the Galactic disc \citep{friel1995old,anders2014chemodynamics,netopil2016metallicity}. This panel establishes the empirical relation, but does not yet distinguish between uncertainty in the inferred gradient and real astrophysical dispersion about that gradient.

To quantify the mean trend, we perform a Monte Carlo linear regression in which both $R_{\mathrm{g}}$ and $[\mathrm{Fe/H}]$ are perturbed within their measurement uncertainties. For the full SCIENCE sample, this yields a well-constrained gradient of \(\nabla_{R_{\mathrm{g}}} = -0.076 \pm 0.008~\mathrm{dex~kpc^{-1}}\), as shown in the upper-right panel. The narrow shaded band in this panel represents the uncertainty on the mean relation (i.e. the uncertainty in the fitted slope and intercept). Its small width indicates that the ensemble of clusters tightly constrains the average metallicity gradient of the disc.

However, the data exhibit a much broader spread about the mean relation than can be explained by measurement uncertainties alone. We therefore infer an intrinsic scatter term, $\sigma_{\rm int}$, from the residuals by requiring that the reduced $\chi^2$ of the fit is approximately unity, obtaining \(\sigma_{\rm int} = 0.118~\mathrm{dex}\). The wide shaded region shown in the upper-right panel is the corresponding prediction band, which combines the uncertainty in the mean relation with this intrinsic scatter. This band represents the expected cluster-to-cluster metallicity spread at fixed $R_{\mathrm{g}}$. Consequently, the fact that many clusters lie well outside the narrow mean-relation band is not anomalous; instead, it demonstrates that the observed dispersion is dominated by real astrophysical scatter rather than by uncertainty in the gradient itself.

To assess whether this behaviour depends on age, we split the SCIENCE sample into two conservative age bins, young ($\tau < 1$~Gyr) and old ($\tau \geq 1$~Gyr). A two-sample Kolmogorov--Smirnov test comparing the metallicity distributions of these subsamples yields $D = 0.667$ and $p = 1.6\times10^{-5}$, indicating that the young and old clusters are not drawn from the same parent $[\mathrm{Fe/H}]$ distribution. The lower-left panel shows the $[\mathrm{Fe/H}]$--$R_{\mathrm{g}}$ relation for the younger clusters. A weighted linear regression yields a slope of \(\nabla_{R_{\mathrm{g}}} = -0.048 \pm 0.005~\mathrm{dex~kpc^{-1}}\),
shallower than the global relation. The younger clusters define a relatively tight metallicity sequence with a shallower radial gradient than the older population.

In contrast, the lower-right panel shows the older population ($\tau \geq 1$ Gyr), which exhibits a steeper metallicity gradient, $\nabla_{R_{\rm g}} = -0.078 \pm 0.017~{\rm dex~kpc^{-1}}$, than the younger population, $\nabla_{R_{\rm g}} = -0.048 \pm 0.005~{\rm dex~kpc^{-1}}$. Although this difference should be interpreted cautiously given the larger scatter and relatively small sample size of the older bin, these clusters exhibit both a steeper gradient and substantially larger dispersion about the best-fitting relation. This behaviour may reflect a combination of secular radial redistribution, age-dependent survival effects, and chemically distinct outer-disc populations among the older clusters \citep{sellwood2002radial,minchev2013chemodynamical,lamers2005disruption,magrini2023gaia}. The larger scatter in the older population is driven, in part, by a small number of metal-poor clusters with [Fe/H] $\lesssim -0.4$. To investigate whether this difference could arise from a systematic abundance-scale offset relative to the Gaia--ESO gradient analysis of \cite{magrini2023gaia}, we cross-matched our final sample with their open-cluster catalogue. The two samples have 41 clusters in common. For these clusters, the median abundance offsets, defined as this work minus \cite{magrini2023gaia}, are $-0.05$ dex in [Fe/H], $+0.02$ dex in [Mg/H], and $+0.06$ dex in [Mg/Fe], with dispersions of 0.07, 0.05, and 0.08 dex, respectively. Overall, the abundance scales of the common clusters are broadly consistent.

The difference between the global distributions is therefore mainly due to the different metallicity coverage of the two samples rather than to the use of $R_\mathrm{g}$. In particular, our final sample contains more low-metallicity clusters, with five clusters at [Fe/H] $< -0.4$, whereas \cite{magrini2023gaia} include only one such cluster. Conversely, \cite{magrini2023gaia} include more metal-rich clusters, with eight clusters at [Fe/H] $> +0.2$ compared with two in our sample. We therefore caution against over-interpreting the detailed shape of the outer-disc metallicity distribution in the present sample.

Across all panels, the GOLD subsample closely follows the same global locus as the full SCIENCE sample and reproduces both the mean gradient and the intrinsic dispersion. This confirms that the measured trends are not driven by a small number of poorly constrained systems, but instead reflect the underlying chemo-dynamical structure of the Galactic open-cluster population.

\subsection{Alpha-element chemistry and radial mixing in the Galactic disc}

\begin{figure}
 \includegraphics[width=\columnwidth]{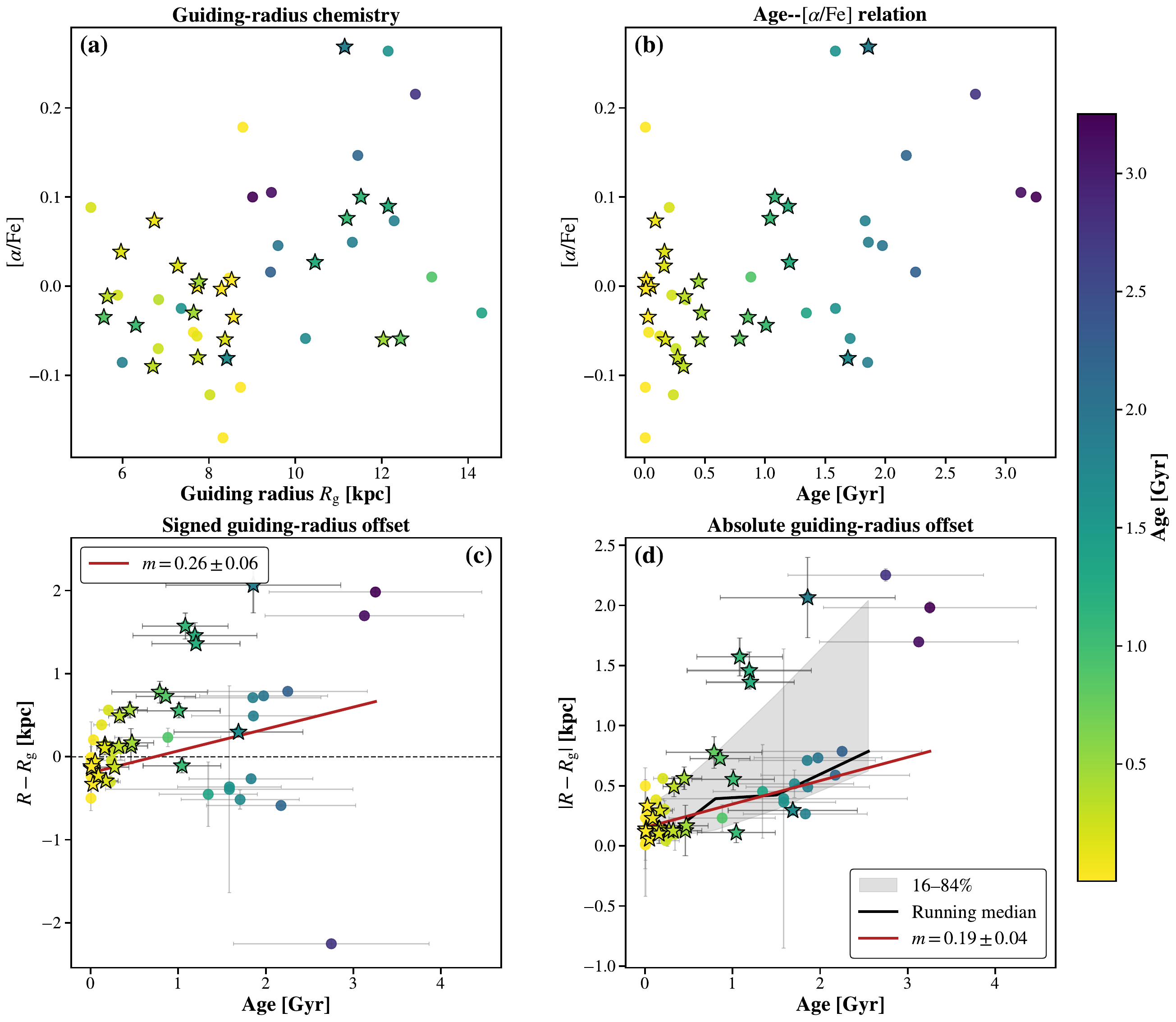}
 \caption{
Chemo-dynamical trends of the Gaia--ESO open-cluster sample. 
(a) Mean $[\alpha/\mathrm{Fe}]$ abundance as a function of guiding radius, $R_{\mathrm{g}}$. 
(b) Mean $[\alpha/\mathrm{Fe}]$ versus cluster age. 
(c) Signed radial offset, $R-R_{\mathrm{g}}$, as a function of age. 
(d) Absolute radial offset, $|R-R_{\mathrm{g}}|$, as a function of age. 
Blue circles and orange stars denote the SCIENCE and GOLD samples, respectively. 
Solid lines show weighted linear fits to the SCIENCE sample, while black squares indicate binned medians with $16$--$84\%$ percentile ranges. 
The lower panels illustrate the increasing spread of non-circular orbital structure with age.
}
\label{fig:5}
\end{figure}

Figure~\ref{fig:5} presents the chemo-dynamical trends of the Gaia--ESO open cluster sample in guiding-radius, age, and migration-sensitive parameter space.
Panel (a) shows the mean $[\alpha/\mathrm{Fe}]$ abundance, computed from Mg, Si, and Ca, as a function of guiding radius, $R_{\mathrm{g}}$.
The distribution exhibits substantial scatter, although a weak tendency toward enhanced $[\alpha/\mathrm{Fe}]$ at larger $R_{\mathrm{g}}$ is visible.
Clusters at small guiding radii are predominantly young and cluster around solar $[\alpha/\mathrm{Fe}]$, whereas moderately enhanced $[\alpha/\mathrm{Fe}]$ values are preferentially associated with clusters at larger $R_{\mathrm{g}}$.

At face value, the present-day distribution does not map trivially onto the simplest expectations from inside--out disc formation models, in which the inner disc undergoes more rapid early enrichment \citep{matteucci1989galactic,chiappini1997chemical}. However, such expectations apply most directly when present-day radius closely traces birth radius, an assumption that may not hold for dynamically evolving cluster populations.
For open clusters, the observed $[\alpha/\mathrm{Fe}]$--$R_{\mathrm{g}}$ distribution likely reflects a combination of age-dependent chemical evolution, orbital redistribution, preferential survival of clusters at larger radii, and selection effects, rather than a purely intrinsic radial increase in $[\alpha/\mathrm{Fe}]$.

The large dispersion at fixed $R_{\mathrm{g}}$ indicates that present-day angular momentum alone does not uniquely determine a cluster's chemical properties. This interpretation is clarified by the age dependence shown in Panel~(b).
Older clusters systematically exhibit modestly higher $[\alpha/\mathrm{Fe}]$ ratios, qualitatively consistent with the time-dependent chemical evolution of the Galactic disc \citep[e.g.][]{tinsley1979stellar,matteucci1986relative}. Since the analysed clusters belong predominantly to the low-$[\alpha/\mathrm{Fe}]$ disc regime, the observed trend should not be interpreted as evidence that the older clusters formed prior to the onset of significant Type~Ia supernova enrichment. Instead, the behaviour likely reflects the continued iron enrichment of the interstellar medium by Type~Ia supernovae over time, which gradually lowers the mean $[\alpha/\mathrm{Fe}]$ abundance in younger stellar populations.

To disentangle the relative roles of age and radius, we performed a weighted multivariate regression of the form
\begin{equation}
[\alpha/\mathrm{Fe}] = a + b\,R_{\mathrm{g}} + c\,\mathrm{Age}.
\end{equation}

The regression demonstrates that age is the dominant predictor of $[\alpha/\mathrm{Fe}]$ ($c = 0.038 \pm 0.015~\mathrm{dex\,Gyr^{-1}}$, $p = 0.017$). Although a weak positive trend is visible in Panel~(a), the residual dependence on guiding radius at fixed age is statistically insignificant ($b = 0.010 \pm 0.006~\mathrm{dex\,kpc^{-1}}$, $p = 0.112$). The model explains a modest fraction of the variance in $[\alpha/\mathrm{Fe}]$ ($R^2 = 0.26$, adjusted $R^2 = 0.23$). Here $R^2$ denotes the coefficient of determination, measuring the fraction of variance in the dependent variable explained by the multivariate linear model. This suggests that the apparent radial behaviour in Panel~(a) is driven primarily by the underlying age distribution of the sample rather than by a strong independent radial dependence.

We note, however, that the age distribution of the cluster sample varies with guiding radius. Older clusters are preferentially found at larger radii, while younger clusters dominate the inner and intermediate disc regions. Part of the observed radial abundance behaviour may therefore arise from this underlying age--radius covariance rather than from purely spatial chemical gradients alone.

The dynamical origin of this chemical dispersion is illustrated in the lower panels of Figure~\ref{fig:5}. Panel~(c) shows the signed offset $R - R_{\mathrm{g}}$ as a function of cluster age. The distribution remains approximately symmetric about zero at all ages, consistent with phase-dependent epicyclic excursions around the guiding radius rather than systematic inward or outward drift.

A weighted linear fit yields a positive slope of $0.264 \pm 0.057~\mathrm{kpc\,Gyr^{-1}}$, consistent with the broader radial excursions associated with the larger orbital eccentricities of older clusters.
Panel~(d) presents the absolute offset $|R - R_{\mathrm{g}}|$, which measures the instantaneous displacement of a cluster from its guiding radius. Because this quantity depends strongly on orbital phase and eccentricity, we use it only as a descriptive measure of non-circular orbital structure rather than as a direct tracer of radial migration or heating.
The typical magnitude of $|R - R_g|$ increases systematically with age, with an empirical slope of
$0.195 \pm 0.036~\mathrm{kpc\,Gyr^{-1}}$ across the sampled age range.
The intrinsic scatter increases from $\sigma \approx 0.15$~kpc in the youngest age bins to $\sigma \approx 0.71$~kpc in the oldest bins, quantitatively confirming the broader distribution observed in Panels~(c) and (d).
Within an axisymmetric framework, this behaviour is qualitatively consistent with progressively larger epicyclic excursions in older clusters. However, the observed broadening likely reflects a combination of secular dynamical evolution, preferential survival of dynamically hotter clusters, and enhanced disruption in the inner Galactic disc.
Because $|R - R_g|$ primarily traces radial blurring rather than net angular-momentum exchange, we do not attempt to isolate churning, which would require explicitly time-dependent non-axisymmetric modelling.

\subsection{Orbital action space}
\begin{figure}
 \includegraphics[width=\columnwidth]{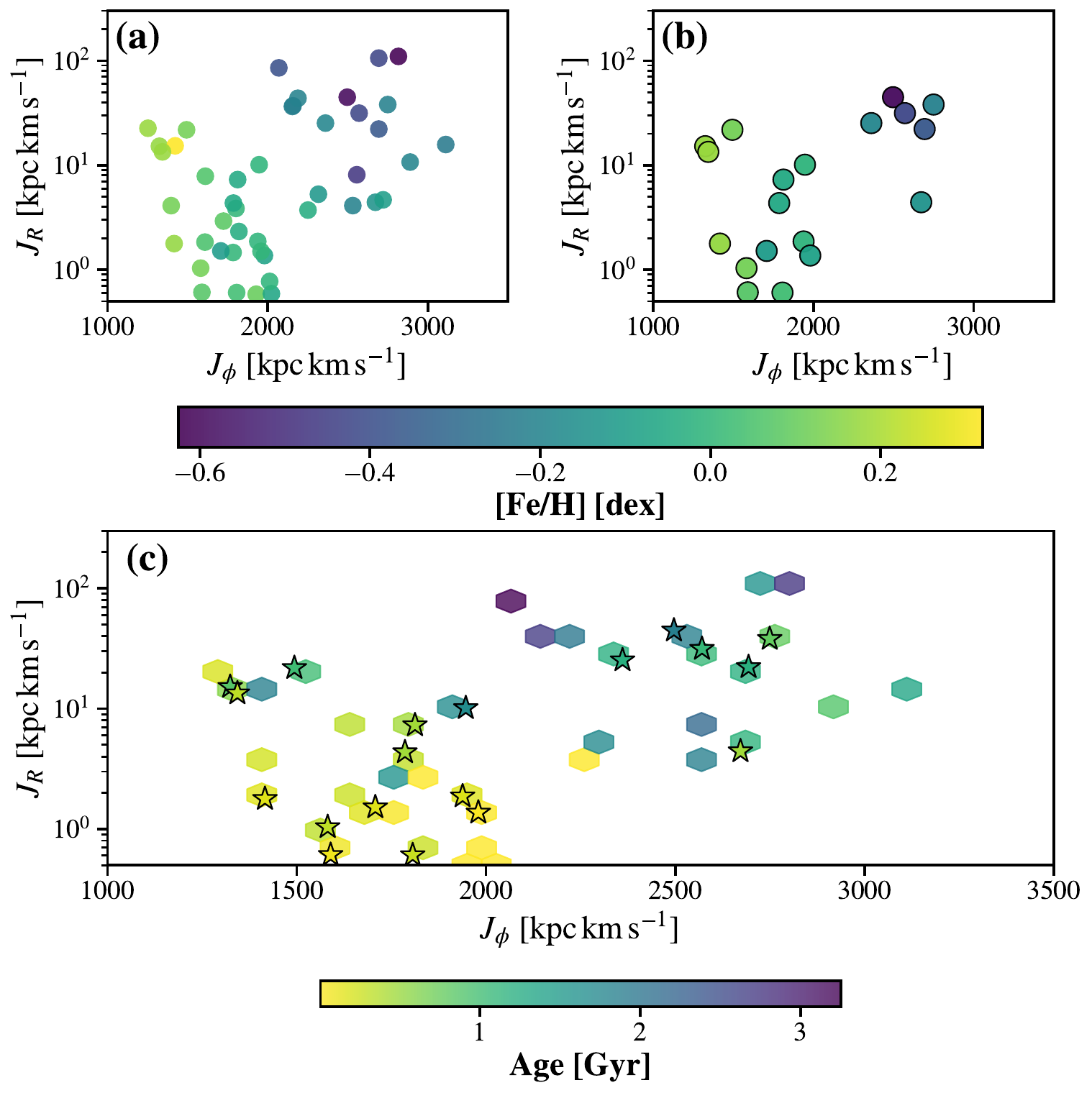}
\caption{
Orbital eccentricity and vertical structure of the Gaia--ESO open-cluster sample as a function of age and guiding radius. 
(a) Orbital eccentricity versus cluster age. 
(b) Eccentricity as a function of guiding radius, $R_{\mathrm{g}}$. 
(c) Maximum vertical excursion, $Z_{\max}$, versus guiding radius. 
(d) Vertical action, $J_Z$, versus cluster age. 
Blue circles and orange stars represent the SCIENCE and GOLD samples, respectively. 
Solid lines show weighted linear fits to the SCIENCE sample, and black squares indicate binned medians with $16$--$84\%$ percentile ranges. 
Older clusters generally exhibit broader orbital distributions and larger vertical amplitudes.
}
\label{fig:6}
\end{figure}

Figure~\ref{fig:6} shows the distribution of open clusters in action space $(J_{\phi}, J_R)$, providing a phase-independent description of their orbital structure \citep{binney2011galactic}. Because actions are adiabatic invariants in time-independent potentials, they offer a dynamically robust framework for identifying orbital families within the Galactic disc. The distribution reveals clear age-dependent stratification. Young clusters (age $<1$~Gyr) occupy dynamically cold, near-circular orbits with low radial action and a mean eccentricity of $\langle e \rangle = 0.051 \pm 0.007$, while clusters older than $2$~Gyr display a broader spread toward higher $J_R$ and larger eccentricities ($\langle e \rangle = 0.173 \pm 0.032$). In an axisymmetric framework, this behaviour reflects cumulative orbital broadening and increased non-circular motion, qualitatively consistent with long-term secular evolution of the Galactic disc \citep{wielen1977diffusion,sellwood2002radial,minchev2013chemodynamical}. %radial blurring (dynamical heating), consistent with expectations for secular disc evolution \citep{wielen1977diffusion,sellwood2002radial,minchev2013chemodynamical}. 
Importantly, the organisation in action space is not purely dynamical. The distribution in angular momentum also exhibits clear chemical structure: near-solar metallicity clusters concentrate around $J_{\phi} \simeq 1800$--$2000~\mathrm{kpc\,km\,s^{-1}}$, whereas older and more metal-poor clusters extend toward higher angular momentum. This coupling between chemistry and orbital configuration is explored in detail in Appendix~\ref{AppenxA} (Figure~\ref{fig:FigA}), where we demonstrate that metallicity gradients, radial action, and chemical residuals all map coherently onto action space. Together, Figures~\ref{fig:6} and \ref{fig:FigA} show that the Galactic disc is stratified simultaneously in age, chemistry, and orbital action, providing direct empirical evidence that secular dynamical evolution reshapes, but does not erase the disc’s chemical structure.

\subsection{Vertical Heating and Dynamical Stability}
\begin{figure}
 \includegraphics[width=\columnwidth]{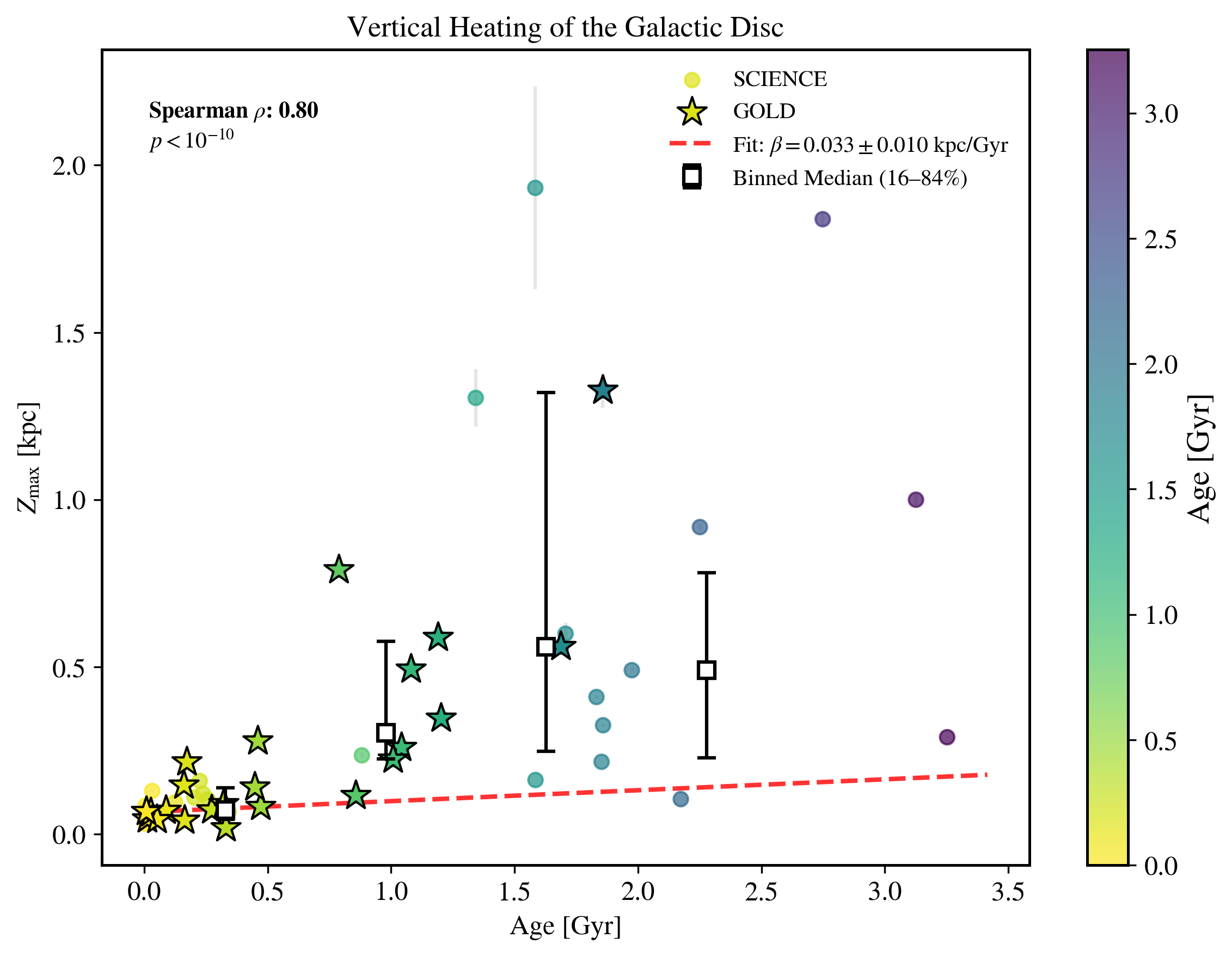}
 \caption{
Maximum vertical orbital excursion, $Z_{\max}$, as a function of cluster age for the Gaia--ESO open-cluster sample. 
Blue circles and orange stars denote the SCIENCE and GOLD samples, respectively. 
The solid line shows the weighted linear fit to the SCIENCE sample, while black squares indicate binned medians with $16$--$84\%$ percentile ranges. 
The Spearman rank coefficient is $\rho = 0.80$ with $p < 10^{-10}$. 
Older clusters preferentially occupy orbits reaching larger heights above the Galactic plane.
}
 
\label{fig:7}
\end{figure}

%Figure~\ref{fig:7} illustrates the vertical dynamical heating of the Galactic disc by showing the maximum vertical excursion, $Z_{\max}$, as a function of cluster age.
Figure~\ref{fig:7} illustrates the age dependence of the vertical orbital structure of the Galactic disc by showing the maximum vertical excursion, $Z_{\max}$, as a function of cluster age. We find a robust, statistically significant monotonic correlation between these quantities, with a Spearman rank coefficient of $\rho = 0.80$ ($p < 10^{-10}$). To ensure that this trend is representative of the bulk population and not driven by a small number of dynamically extreme objects, we performed a leave--out sensitivity analysis (see Appendix~\ref{app:robustness}). The correlation coefficient remains remarkably stable ($\rho \ge 0.78$) even after the successive removal of the five clusters with the largest vertical excursions. This indicates that the observed increase in vertical orbital extent is a global characteristic of the sampled Galactic disc population rather than an effect dominated by a small number of outliers. %This demonstrates that the observed vertical heating is a global characteristic of the sampled Galactic disc population rather than an effect dominated by outliers.
This high correlation coefficient indicates that older clusters preferentially occupy orbits reaching larger heights above the Galactic plane. The observed trend is qualitatively consistent with long-term dynamical evolution of the Galactic disc, although survival bias and selection effects may also contribute to the observed distribution. %This high correlation coefficient confirms that older clusters preferentially occupy ``hotter'' vertical orbits, reaching larger heights above the Galactic plane as a result of cumulative dynamical scattering.
A weighted linear regression to the SCIENCE sample yields an empirical age–$Z_{\max}$ slope of $0.033 \pm 0.010~\mathrm{kpc\,Gyr^{-1}}$. While this value should not be interpreted as a physical heating ``rate'' in a time-dependent sense, it quantifies the systematic increase in typical vertical orbital amplitude over the sampled age range. Because the open-cluster sample is dominated numerically by younger systems, the fitted relations may be influenced by the uneven age distribution of the sample. We therefore interpret the inferred differences between radial and vertical heating trends cautiously. The shallower vertical trend, compared to the in-plane eccentricity and radial-excursion relations, is consistent with the expected anisotropy of secular disc heating, in which vertical scattering by giant molecular clouds is generally less efficient than radial heating driven by non-axisymmetric structures \citep[e.g.][]{spitzer1953possible,wielen1977diffusion,mackereth2019dynamical}. To verify that the observed trend is not driven solely by the radial dependence of the Galactic potential, we additionally examine the vertical action $J_Z$ in Appendix~\ref{AppenxA}, where we find a nearly identical correlation with age (Figure~\ref{fig:FigA2}). Because $J_Z$ is less sensitive than $Z_{\max}$ to variations in the Galactic potential with radius, we treat it as the more robust dynamical diagnostic, while retaining $Z_{\max}$ here as an intuitive orbital quantity commonly reported in open-cluster orbital studies \citep[e.g.][]{tarricq20213d}. %To confirm that this trend reflects a fundamental increase in orbital energy, we verify in Appendix~\ref{AppenxA} that the vertical action $J_Z$ follows a nearly identical correlation with age (Figure~\ref{fig:FigA2}). 
However, given the significant dispersion in $Z_{\max}$ at all ages, we utilize non-parametric trend indicators to better characterize the population. The binned medians and $16-84\%$ percentile ranges (black squares) reveal a steady increase in the typical vertical extent of the cluster system while highlighting the broad distribution of vertical energies present even among the youngest populations. The consistency between the SCIENCE and GOLD samples (circles and stars, respectively) supports the interpretation that the observed vertical broadening reflects a genuine characteristic of the cluster population rather than arising solely from observational bias. %emphasizes that this vertical heating is an intrinsic physical property of the disc rather than an observational bias.
Finally Table~\ref{tab:final_stats} present summary of the chemo--dynamical and orbital-broadening statistics %the summary of secular heating and migration statistics
derived for the Gaia–ESO open cluster sample.

\section{Discussion}\label{discusion}

Open clusters provide age-resolved, chemically homogeneous tracers of Galactic disc evolution, enabling a direct connection between stellar chemistry, age, and orbital dynamics \citep{friel1995old,anders2014chemodynamics,netopil2016metallicity}. By analysing the \emph{Gaia}--ESO open cluster sample in guiding-radius and action space, we identify clear empirical age-dependent dynamical trends that place cluster populations within the broader framework of secular disc evolution.

The negative $[\mathrm{Fe/H}]$--$R_g$ relation confirms the established radial metallicity gradient of the Galactic disc traced by open clusters \citep{friel1995old,anders2014chemodynamics,netopil2016metallicity}. While the mean gradient is statistically well constrained, the residual dispersion is dominated by intrinsic cluster-to-cluster scatter, which increases toward older ages. This behaviour is consistent with cumulative secular radial mixing operating over Gyr timescales, as predicted in chemo-dynamical models of disc evolution \citep{sellwood2002radial,schonrich2009chemical,minchev2013chemodynamical,frankel2018measuring}. Although the guiding radius provides a phase-independent angular-momentum coordinate \citep{binney2011galactic}, it should not be interpreted as a direct measurement of birth radius; rather, it offers a dynamically meaningful framework for comparing chemistry and orbital structure.

Independent evidence for in-plane heating is provided by the systematic increase of orbital eccentricity and radial excursion amplitude with age. In an axisymmetric potential, this behaviour primarily reflects growth in epicyclic amplitude (radial blurring) \citep{sellwood2002radial,binney2011galactic}. While angular-momentum redistribution (churning) may also contribute \citep{sellwood2002radial,schonrich2009chemical}, isolating that component requires explicitly time-dependent non-axisymmetric modelling \citep{minchev2010new,minchev2013chemodynamical}, which lies beyond the scope of the present study. The coexistence of dynamically hotter and relatively cold clusters at similar ages highlights the stochastic character of secular disc evolution \citep{wielen1977diffusion,sellwood2014secular} and naturally explains the substantial intrinsic scatter observed in the kinematic relations.

The action-space distribution reinforces this interpretation. Actions are adiabatic invariants in smooth potentials and provide a phase-independent description of orbital families \citep{binney2011galactic,BinneyTremaine2008}. Clusters occupy a well-defined low-$J_R$ locus characteristic of disc orbits, while older and more metal-poor systems extend to higher radial actions, consistent with progressive dynamical heating. At the same time, clusters with similar chemical properties remain grouped in coherent regions of $(J_R, J_\phi)$ space despite substantial differences in present-day radius. This behaviour is consistent with radial migration redistributing angular momentum while largely preserving chemical coherence, in agreement with chemo-dynamical models \citep{minchev2013chemodynamical,frankel2018measuring,sanders2014actions}.

Vertical trends provide complementary evidence for secular heating. The strong monotonic correlation between age and maximum vertical excursion indicates that open clusters participate in the gradual heating processes inferred for field-star populations \citep{aumer2009kinematics,mackereth2019dynamical}. The absence of a strong dependence of $Z_{\max}$ on guiding radius suggests that vertical heating operates across a broad range of disc radii, consistent with predominantly local scattering processes such as interactions with giant molecular clouds \citep{spitzer1953possible,wielen1977diffusion}. The corresponding age dependence of the vertical action $J_Z$ confirms that the observed relation reflects a systematic increase in vertical orbital energy \citep{BinneyTremaine2008}.

The $\alpha$-element trends further illustrate the coupling between chemistry and age. The enhancement of $[\alpha/\mathrm{Fe}]$ in older clusters is qualitatively consistent with expectations from inside--out disc formation \citep{matteucci1989galactic,chiappini1997chemical} and the temporal transition from core-collapse to Type~Ia supernova enrichment \citep{tinsley1979stellar,matteucci1986relative}. When examined in action space, these chemical signatures remain coherent despite orbital redistribution, underscoring the value of clusters as clean tracers of the disc’s chemo-dynamical evolution.

Most previous open-cluster studies have concentrated on radial abundance gradients and their temporal variation \citep[e.g.][]{friel2002metallicities,anders2017red,donor2020open}, typically using present-day positions. By explicitly combining homogeneous chemistry with orbital actions and guiding radii, the present work situates open clusters within the chemo-dynamical framework now widely applied to field-star samples \citep{mackereth2019dynamical,trick2019galactic,frankel2020keeping}, providing an age-calibrated empirical benchmark for disc evolution studies.
\subsection{Limitations}

The primary limitation of this analysis is the modest number of clusters with both high-quality abundances and full six-dimensional phase-space information. Selection effects, including spectroscopic targeting and the under-representation of highly extincted inner-disc regions, may influence the detailed form of some relations. Nevertheless, the close agreement between the SCIENCE and GOLD subsamples indicates that the principal trends identified here are not driven by low-quality measurements.

The orbital calculations adopt a static, axisymmetric Galactic potential. In reality, non-axisymmetric and time-dependent perturbations such as the Galactic bar and transient spiral structure are expected to contribute to radial migration and disc heating \citep{sellwood2002radial,minchev2010new,minchev2013chemodynamical}. While such perturbations may affect the detailed orbital histories of individual clusters, the ensemble trends identified here, particularly the persistence of a negative metallicity gradient combined with increasing dispersion at older ages, are consistent with secular evolution operating over Gyr timescales.

Future \emph{Gaia} data releases, together with ongoing and next-generation spectroscopic surveys, will substantially expand the open-cluster sample in both spatial coverage and chemical diversity. In particular, the VVV--GALCEN programme is designed to target a large and systematically selected cluster sample spanning the inner disc including heavily obscured regions along bulge lines of sight  out to the outer Galactic disc, and covering a broad range of ages and metallicities. By combining near-infrared spectroscopy with deep VVVX photometry, this survey will probe cluster populations that are currently under-represented in optical studies, especially in high-extinction environments. 

Such expanded radial and vertical coverage will enable significantly tighter constraints on the efficiency of radial migration and secular disc heating, and will allow the empirical framework developed here to be extended to larger samples and, ultimately, to time-dependent Galactic potentials. Open clusters therefore provide a robust observational foundation for anchoring models of disc formation and long-term secular evolution to well-calibrated age--chemistry--dynamics benchmarks.

\section{Summary and conclusions}\label{summary}

We have presented a systematic chemo-dynamical analysis of open clusters from the \emph{Gaia}--ESO Survey, combining homogeneous cluster chemistry with precise ages and full six-dimensional phase-space information from \emph{Gaia} DR3. By analysing the sample in guiding-radius and action space, we tested whether clusters with similar chemical properties occupy coherent orbital families and whether their ensemble behaviour encodes the signatures of radial migration and secular disc heating. Our results demonstrate that open clusters provide a high-fidelity, age-resolved record of the chemo-dynamical stratification of the Galactic disc. The main conclusions are as follows:

\begin{enumerate}
    \item We confirm a statistically significant negative radial metallicity gradient ($\nabla_{R_{\mathrm{g}}} = -0.076 \pm 0.008~\mathrm{dex\,kpc^{-1}}$). 
The scatter about the mean relation is dominated by intrinsic cluster-to-cluster dispersion ($\sigma_{\rm int} \approx 0.12$~dex), which increases with age, consistent with cumulative secular radial mixing.

  \item We quantify the growth of in-plane orbital excursions with age using $|R - R_g|$ as a proxy for radial blurring. The excursion amplitude increases systematically with age, with an empirical slope of $0.195 \pm 0.036~\mathrm{kpc\,Gyr^{-1}}$, indicating progressively larger epicyclic amplitudes and dynamically hotter in-plane structure in older cluster populations.
  
    \item We identify a statistically significant age--eccentricity trend, with a strong monotonic correlation between age and eccentricity (Spearman $\rho = 0.72$, $p = 3.1 \times 10^{-9}$). An orthogonal-distance regression fit yields an empirical slope of $m = 0.081 \pm 0.013~\mathrm{Gyr^{-1}}$, with a robust Theil--Sen estimate of $m \simeq 0.056~\mathrm{Gyr^{-1}}$. This behaviour indicates progressive broadening of orbital eccentricities with age, consistent with cumulative in-plane dynamical heating. The residual variance of the fit ($\simeq 1.3$) suggests that, while the global trend is well defined, individual clusters likely experience stochastic dynamical histories.

   \item We observe a strong age dependence of vertical orbital amplitude, with an empirical age–$Z_{\max}$ slope of $0.033 \pm 0.010~\mathrm{kpc\,Gyr^{-1}}$. Compared to the in-plane trends, this shallower relation indicates that vertical excursions increase more slowly with age than radial excursions over the sampled range. This relative difference is consistent with the expected anisotropy of secular disc heating, in which in-plane perturbations and epicyclic growth are generally more efficient than vertical thickening driven by local scattering processes \citep[e.g.][]{spitzer1953possible,wielen1977diffusion,mackereth2019dynamical}.
 
    \item We find empirical evidence for the coupling of chemical and dynamical evolution. The significant correlation between radial action and metallicity residuals ($r_s = 0.378, p = 0.0068$) is consistent with a scenario in which dynamical heating and chemical dispersion are coupled through secular processes.

    \item Our action-space analysis demonstrates that the Galactic disc is simultaneously stratified in age, chemistry, and orbital action. Clusters with similar chemical properties occupy coherent regions in $(J_R, J_\phi)$ space despite significant radial redistribution. This indicates that secular evolution reshapes angular momentum while largely preserving chemical coherence, establishing open clusters as high-fidelity tracers of disc chemo-dynamical history.

\end{enumerate}
Taken together, these results show that open clusters provide a uniquely powerful analogue to field-star studies, with the critical advantage of precise ages. By combining guiding radii and orbital actions with homogeneous chemistry, we demonstrate that the Galactic disc is stratified coherently in age, metallicity, and dynamical phase space. This work therefore provides empirical insight into the coupled evolution of cluster chemistry, orbital structure, and secular disc heating over the last $\sim4$~Gyr, and establishes open clusters as benchmark tracers for anchoring models of Galactic secular evolution. 

While the present study is limited by sample size, forthcoming \emph{Gaia} data releases and next-generation spectroscopic surveys (VVV-GALCEN, WEAVE, and 4MOST) will enable increasingly stringent tests. Extending this framework to time-dependent Galactic potentials will firmly anchor models of disc formation to observational benchmarks.
%%%%%%%%%%%%%%%%%%%%%%%%%%%%%%%%%%%%%%%%%%%%%%%%%%
\section*{Data availability}
The data used in this article are publicly available from the ESO Science Archive Facility. We used the final public data release of the \emph{Gaia}--ESO Survey (2023), which can be accessed at \url{https://archive.eso.org}. The derived data products generated in this work are presented in Tables~\ref{tab:orbital_parameters_science} and \ref{quality_cut_fig}. The full versions of Tables~2 and B.1 are available in electronic form at the CDS via anonymous ftp to \texttt{cdsarc.u-strasbg.fr} (130.79.128.5) or via \url{http://cdsweb.u-strasbg.fr/cgi-bin/qcat?J/A+A/}.

\begin{acknowledgements} 
This work was funded by the Postdoctoral Talent Attraction Competition for Research Centers and Institutes of the Universidad Andrés Bello (UNAB) 2025, project Nº. DI-07-25/ATP.  M.G. gratefully acknowledges support from Fondecyt through grant 1240755. J.G.F-T gratefully acknowledges the support provided by ANID Fondecyt Regular No. 1260371, the Joint Committee ESO-Government of Chile under the agreement 2023 ORP 062/2023 and the support of the Doctoral Program in Artificial Intelligence, DISC-UCN. D.M. gratefully acknowledges support from the Center for Astrophysics and Associated Technologies CATA by the ANID BASAL projects ACE210002 and FB210003, by Fondecyt Project No. 1220724. B.B. akcnowledges partial financial suppoert from FAPESP, CNPq, and CAPES. BD acknowledges support from ANID Basal project FB210003. Finally, we thank the anonymous referee for the careful reading of the manuscript and for the constructive comments and suggestions, which significantly improved the clarity and interpretation of this work.

This work is based on data products from observations made with ESO Telescopes at the La Silla Paranal Observatory under programme ID 188.B-3002.  The Gaia–ESO Survey is a public spectroscopic survey conducted with the FLAMES instrument on the Very Large Telescope at the ESO Paranal Observatory.  The Survey has been supported by the European Union through ERC grant number 320360 and by the national institutions participating in the Gaia–ESO Consortium. We acknowledge the work of the Gaia–ESO Survey consortium in providing high-quality spectroscopic data products.
\end{acknowledgements}

%%%%%%%%%%%%%%%%%%%%%%%%%%%%%%%%%%%%%%%%%%%%%%%%%%%%%%%%%%%%%%
%%%%%%%%%%%%%%%%%%%%%%%%%%%%%%%%%%%%%%%%%%%%%%%%%%%%%%%%%%%%%%
\bibliographystyle{aa}
\bibliography{Ref}

\FloatBarrier %\usepackage{placeins}
\clearpage
\begin{appendix}
\nolinenumbers
\section{Robustness of the Age--\texorpdfstring{$Z_{\max}$}{Zmax} Correlation}
\label{app:robustness}

To verify that the observed age--$Z_{\max}$ correlation represents a global property of the cluster population rather than the influence of a few dynamically ``hot'' outliers, we conducted a sensitivity analysis. We recomputed the Spearman rank correlation coefficient ($\rho$) by progressively excluding clusters with the highest $Z_{\max}$ values from the SCIENCE sample.

As shown in Table~\ref{tab:robust_spearman}, the correlation remains both strong and highly significant. The coefficient decreases only marginally, from $\rho = 0.80$ ($N=50$) to $\rho = 0.78$ ($N=45$) after removing the five most extreme objects. The $p$-values remain consistently below $10^{-9}$, allowing us to reject the null hypothesis of no correlation with high confidence in all cases. This stability demonstrates that the vertical heating trend is an intrinsic feature of the disc's secular evolution, characterized by a steady increase in vertical excursion across the entire age range studied.

\section{Additional action-space projections}
\begin{figure}[!t]
 \includegraphics[width=\columnwidth]{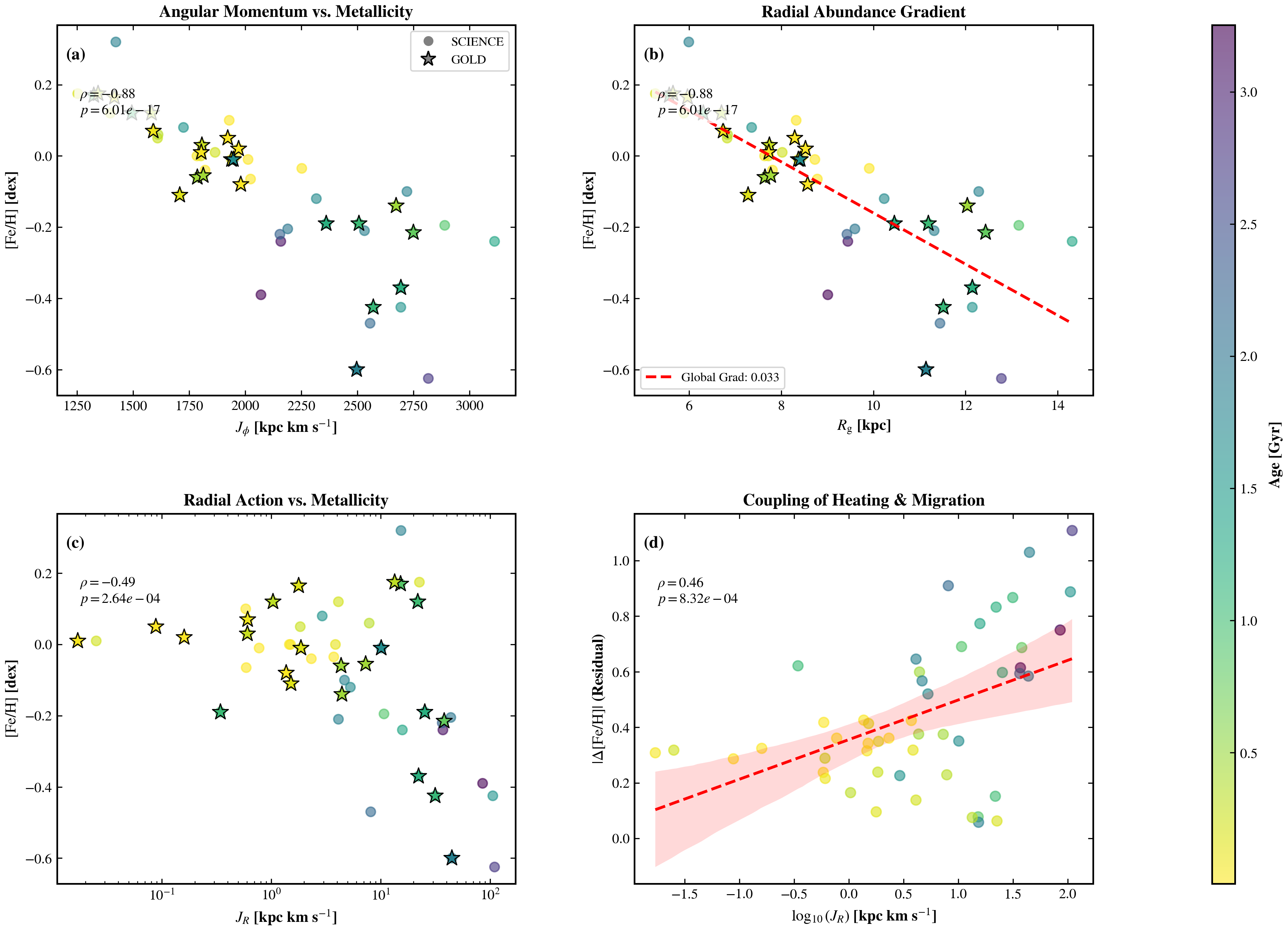}
\caption{
Mean cluster metallicity, $[\mathrm{Fe/H}]$, as a function of three complementary orbital quantities:
azimuthal action $J_{\phi}$ (left), guiding radius $R_{\mathrm{g}}$ (centre), and radial action $J_{R}$ (right).
Points are colour-coded by cluster age, with the SCIENCE sample shown as circles and the higher-quality GOLD subsample highlighted by star symbols.
The $J_{\phi}$ projection emphasises the tight connection between chemical enrichment and angular momentum,
while the $R_{\mathrm{g}}$ panel provides an intuitive reference to the radial metallicity gradient.
The $J_{R}$ projection illustrates the extension of older and more metal-poor clusters toward dynamically hotter orbits,
consistent with progressive radial heating.
Together, these panels demonstrate that the observed metallicity structure of the disc is more fundamentally organised in action space than in present-day radius.
}
\label{fig:FigA}
\end{figure}

Figures~\ref{fig:FigA} and \ref{fig:FigA2} present supplementary action-space projections that support the interpretation adopted in the main text. 
Building on the dynamical stratification identified in Figure~\ref{fig:6}, Figure~\ref{fig:FigA} presents complementary chemo-dynamical projections in action space. These diagnostics demonstrate that the age-dependent heating observed in the main text is accompanied by systematic chemical organisation. Panel (a) shows $[\mathrm{Fe/H}]$ as a function of azimuthal action $J_{\phi}$. The monotonic decrease of metallicity with increasing angular momentum reflects the fundamental radial ordering of the Galactic disc in action space. This trend is consistent with the established radial metallicity gradient traced by open clusters \citep[e.g.][]{friel1995old,anders2017galactic,netopil2016metallicity}.
Panel (b) illustrates the $[\mathrm{Fe/H}]$--$R_{\mathrm{g}}$ relation separated by age. Younger clusters ($\lesssim 1$~Gyr) define a relatively tight sequence with slope $-0.052 \pm 0.010$~dex\,kpc$^{-1}$, while older clusters ($\gtrsim 3$~Gyr) show comparable slopes within uncertainties but significantly larger dispersion. The increased scatter at older ages is consistent with cumulative secular radial mixing \citep{sellwood2002radial,minchev2013chemodynamical,anders2017galactic}. Panel (c) connects chemical structure to orbital heating. Solar-metallicity clusters occupy a dynamically cold regime ($J_R \lesssim 5~\mathrm{kpc\,km\,s^{-1}}$), whereas older and more metal-poor clusters extend toward higher radial actions. This behaviour mirrors the age-dependent broadening of eccentricity and radial excursions discussed in the main text. Finally, Panel (d) quantifies the coupling between heating and chemical dispersion. Defining the metallicity residual $|\Delta \mathrm{[Fe/H]}|$ relative to the global $[\mathrm{Fe/H}]$--$R_{\mathrm{g}}$ relation, we find a statistically significant positive correlation with radial action ($r_s = 0.378$, $p = 6.8 \times 10^{-3}$). Clusters on dynamically hotter orbits, therefore, exhibit larger chemical offsets from the mean radial trend. This result provides direct empirical evidence that the same secular processes responsible for increasing orbital non-circularity also promote chemical mixing across the Galactic disc.
\begin{table}
\centering
\caption{Robustness of the Spearman rank correlation between age and $Z_{\max}$ for the SCIENCE sample after excluding the $N_{\mathrm{rem}}$ clusters with the largest vertical excursions.}
\label{tab:robust_spearman}
\begin{tabular}{cccc}
\hline
\hline
$N_{\mathrm{rem}}$ & $\rho$ & $p$-value & $N_{\mathrm{final}}$ \\
\hline
0 & 0.80 & $2.8\times10^{-12}$ & 50 \\
1 & 0.80 & $3.3\times10^{-12}$ & 49 \\
2 & 0.79 & $1.7\times10^{-11}$ & 48 \\
3 & 0.79 & $4.9\times10^{-11}$ & 47 \\
4 & 0.79 & $7.9\times10^{-11}$ & 46 \\
5 & 0.78 & $4.1\times10^{-10}$ & 45 \\
\hline
\end{tabular}
\end{table}
\section{Additional chemo-dynamical diagnostics}\label{AppenxA}

\begin{figure}[!t]
 \includegraphics[width=\columnwidth]{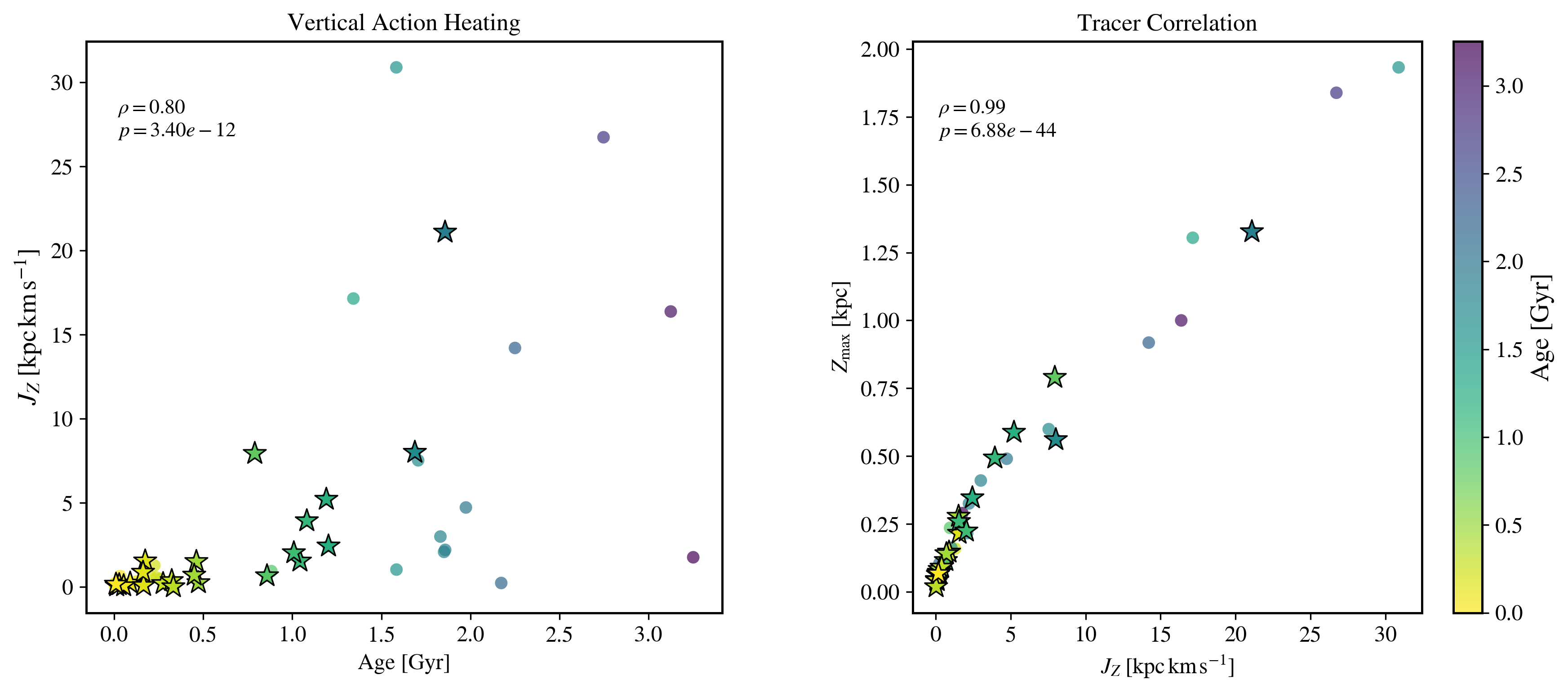}
 \caption{
Vertical dynamical diagnostics for the \emph{Gaia}--ESO open cluster sample.
Left panel: vertical action $J_{Z}$ as a function of cluster age, showing a systematic increase toward older populations,
consistent with progressive vertical heating.
Right panel: maximum vertical excursion $Z_{\max}$ as a function of $J_{Z}$, illustrating the close correspondence between
the phase-independent vertical action and the orbit-based geometric measure.
The tight correlation between $Z_{\max}$ and $J_{Z}$ demonstrates that the age--$Z_{\max}$ relation discussed in the main text
reflects genuine vertical dynamical heating rather than orbital phase effects.
}
\label{fig:FigA2}
\end{figure}

 Figure~\ref{fig:FigA2} provides complementary vertical diagnostics that support the trends discussed in Figure \ref{fig:7}. Panel~(a) shows a strong monotonic increase of the vertical action $J_{Z}$ with cluster age ($\rho = 0.80$), confirming that the progressive vertical ``puffing up'' of the Galactic disc is a fundamental dynamical process. Because $J_Z$ is an adiabatic invariant in smooth, slowly varying potentials \citep{BinneyTremaine2008}, its systematic increase with age provides strong evidence for cumulative vertical heating driven by discrete scattering processes. Panel~(b) illustrates a remarkably tight correlation between the maximum vertical excursion $Z_{\max}$ and the vertical action $J_{Z}$ ($\rho = 0.99$). This near-linear mapping demonstrates that $Z_{\max}$ serves as a robust geometric proxy for the vertical orbital structure across the sampled energy
range. Moreover, it confirms that the age--$Z_{\max}$ relation discussed in Figure \ref{fig:7} reflects genuine vertical dynamical heating, rather than transient orbital phase effects or inconsistencies in the adopted Galactic potential model \citep{BinneyTremaine2008}.

\begin{table*}
\centering
\caption{Summary properties of the final sample of 59 Gaia--ESO open clusters. }
%\caption{Summary properties of the final sample of Gaia--ESO open clusters. Only the first five rows are shown; the full machine-readable table is available electronically.}
\label{quality_cut_fig}
\tiny
\setlength{\tabcolsep}{1pt}
\renewcommand{\arraystretch}{0.75}
\resizebox{\textwidth}{!}{%
\begin{tabular}{lcccccccccccccccccc}
\toprule
ID &
RA &
Dec &
$N_{\rm obs}$ &
$N_{\rm mem}$ &
$\mathrm{SNR}_{\rm med}$ &
$\mathrm{RV}_{\rm mean}$ &
$\mathrm{RV}_{\rm std}$ &
$\mathrm{[Fe/H]}$ &
$\sigma_{\mathrm{[Fe/H]}}$ &
$A(\mathrm{Mg})$&
$\sigma_{A(\mathrm{Mg})}$ &
$\mu_{\alpha*}$ &
$\sigma_{\mu_{\alpha*}}$ &
$\mu_{\delta}$ &
$\sigma_{\mu_{\delta}}$ &
$\varpi$ &
$d$ &
$\log(\mathrm{Age})$ \\
\midrule
Blanco1$^{\mathrm{SG}}$ & 0.914 & -30.010 & 21 & 21 & 182.910 & 5.746 & 2.170 & -0.010 & 0.029 & 7.495 & 0.088 & 18.724 & 0.017 & 2.592 & 0.015 & 4.231 & 234.390 & 8.239 \\
Br22$^{\mathrm{SG}}$ & 89.615 & 7.758 & 18 & 18 & 33.245 & 94.476 & 1.399 & -0.215 & 0.103 & 7.200 & 0.014 & 0.563 & 0.012 & -0.389 & 0.010 & 0.134 & 5334.363 & 8.897 \\
Br30$^{\mathrm{SG}}$ & 104.436 & 3.227 & 11 & 11 & 28.260 & 46.485 & 2.028 & -0.140 & 0.148 & 7.390 & 0.014 & -0.218 & 0.007 & -0.313 & 0.006 & 0.176 & 4462.350 & 8.663 \\
Br31$^{\mathrm{SG}}$ & 104.406 & 8.288 & 49 & 49 & 31.030 & 56.582 & 1.981 & -0.370 & 0.103 & 7.260 & 0.014 & 0.114 & 0.008 & -0.921 & 0.006 & 0.122 & 5864.814 & 9.075 \\
Br36$^{\mathrm{SG}}$ & 109.100 & -13.193 & 104 & 104 & 27.365 & 62.833 & 1.756 & -0.190 & 0.148 & 7.359 & 0.185 & -1.692 & 0.006 & 0.957 & 0.004 & 0.212 & 3839.619 & 9.018 \\
\bottomrule
\end{tabular}%
}
\tablefoot{Columns give mean coordinates, sample sizes, kinematics, chemistry, Gaia DR3 astrometry, distance, and age. Values are truncated to three decimal places. Superscripts indicate sample membership: $^{\mathrm{G}}$ (GOLD) and $^{\mathrm{S}}$ (SCIENCE). Only the first five rows are shown; the full machine-readable table is available electronically.}
\end{table*}

% \clearpage
% \onecolumn
% \begin{landscape}
% \tiny
% \setlength{\tabcolsep}{1pt}
% \renewcommand{\arraystretch}{0.58}

% \input{table_clusters_longB}\label{quality_cut_fig}

% \normalsize
% \end{landscape}
% \twocolumn
\end{appendix}

\end{document}